\documentclass[floatfix,aps,twocolumn,prb,superscriptaddress,amsmath,amssymb]{revtex4-1}
\usepackage{lmodern}
\usepackage{longtable}
\usepackage{bm}
\usepackage[rgb]{xcolor}
\usepackage[author={Pablo S. Cornaglia}]{pdfcomment}
	\usepackage{multirow}
	\usepackage[pdftex]{graphicx}
	\usepackage[utf8]{inputenc}

	\usepackage{tabularx}
\usepackage{array} 
\usepackage[T1]{fontenc}
\newcolumntype{C}[1]{>{\centering\arraybackslash}p{#1}}

\graphicspath{{./figs/}}

\begin{document}
\title{Why the Co-based 115 compounds are different?: The case study of GdMIn$_5$ (M=Co, Rh, Ir)}
\author{Jorge I. Facio}
\affiliation{Centro At{\'o}mico Bariloche and Instituto Balseiro, CNEA, 8400 Bariloche, Argentina}
\affiliation{Consejo Nacional de Investigaciones Cient\'{\i}ficas y T\'ecnicas (CONICET), Argentina}
\author{D. Betancourth}
\affiliation{Centro At{\'o}mico Bariloche and Instituto Balseiro, CNEA, 8400 Bariloche, Argentina}
\affiliation{Consejo Nacional de Investigaciones Cient\'{\i}ficas y T\'ecnicas (CONICET), Argentina}

\author{Pablo Pedrazzini}
\affiliation{Centro At{\'o}mico Bariloche and Instituto Balseiro, CNEA, 8400 Bariloche, Argentina}
\affiliation{Consejo Nacional de Investigaciones Cient\'{\i}ficas y T\'ecnicas (CONICET), Argentina}

\author{V. F. Correa}
\affiliation{Centro At{\'o}mico Bariloche and Instituto Balseiro, CNEA, 8400 Bariloche, Argentina}
\affiliation{Consejo Nacional de Investigaciones Cient\'{\i}ficas y T\'ecnicas (CONICET), Argentina}

\author{V. Vildosola}
\affiliation{Departamento de Materia Condensada, GIyA, CNEA (1650) San Mart\'{\i}n, Provincia de Buenos Aires, Argentina}
\affiliation{Consejo Nacional de Investigaciones Cient\'{\i}ficas y T\'ecnicas (CONICET), Argentina}

\author{D. J. Garc\'{\i}a}
\affiliation{Centro At{\'o}mico Bariloche and Instituto Balseiro, CNEA, 8400 Bariloche, Argentina}
\affiliation{Consejo Nacional de Investigaciones Cient\'{\i}ficas y T\'ecnicas (CONICET), Argentina}

\author{Pablo S. Cornaglia}
\affiliation{Centro At{\'o}mico Bariloche and Instituto Balseiro, CNEA, 8400 Bariloche, Argentina}
\affiliation{Consejo Nacional de Investigaciones Cient\'{\i}ficas y T\'ecnicas (CONICET), Argentina}

\begin{abstract}
The discovery in 2001 of superconductivity in some heavy fermion compounds of the RMIn$_5$ (R=4f or 5f elements, M=Co, Rh, Ir) family, 
has triggered enormous amount of research pointing to understand the physical origin of superconductivity and its relation with magnetism. 
Although many properties have been clarified, there are still crutial questions that remain unanswered. One of these questions is the 
particular role of the transition metal in determining the value of critical superconducting temperature (Tc).
In this work, we analyse an interesting regularity that is experimentally observed in this family of compounds, where the lowest N\'eel 
temperatures are obtained in the Co-based materials. We focus our analysis on the GdMIn$_5$ compounds and perform density-functional-theory 
based total-energy calculations to obtain the parameters for the exchange coupling interactions between the magnetic moments located at 
the Gd$^{3+}$ ions. 
Our calculations indicate that the ground state of the three compounds is a $C$-type antiferromagnet determined by the competition between 
the first- and second-neighbor exchange couplings inside GdIn$_3$ planes and stabilized by the couplings across MIn$_2$ planes. 
We then solve a model with these magnetic interactions using a mean-field approximation and Quantum Monte Carlo simulations.
The results obtained for the calculated N\'eel and Curie-Weiss temperatures, the specific heat and the magnetic susceptibility are in very good 
agreement with the existent experimental data.
Remarkably, we show that the first neighbor interplane exchange coupling in the Co-based material is much smaller than in the Rh and Ir analogues 
due to a more two dimensional behaviour in the former. This result explains the observed lower N\'eel temperature in Co-115 systems and may shed 
light on the fact that the Co-based 115 superconductors present the highest Tc.
\end{abstract}

\pacs{75.50.Ee, 63.20.D-, 71.20.-b, 65.40.De}

\maketitle
\section{Introduction}
\begin{figure}[t]
    \begin{center}
        \includegraphics[width=0.3\textwidth]{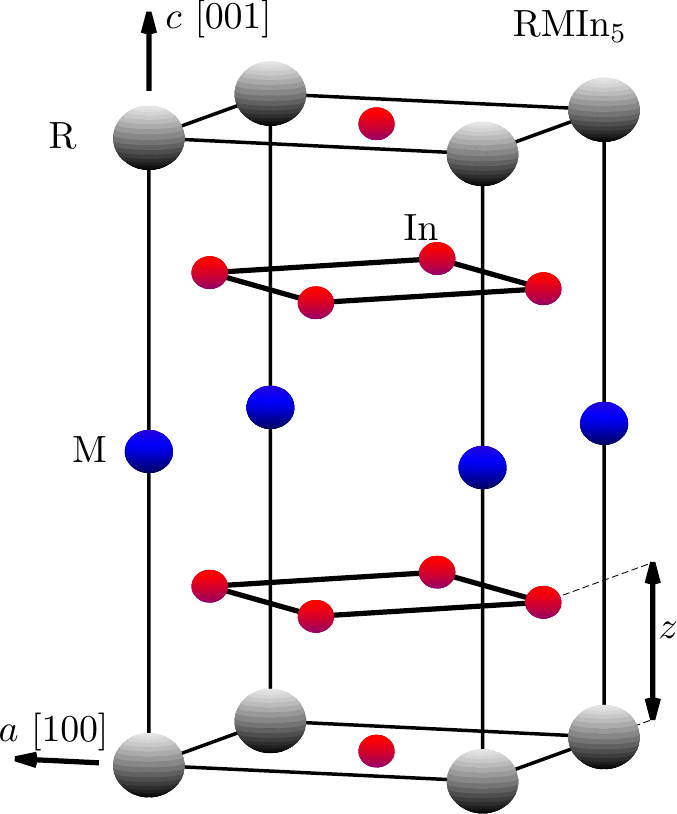}
    \end{center}
    \caption{(Color online) Crystal structure for the RMIn$_5$ compounds. The In atoms are represented by red spheres, the transition metal by blue spheres and the rare earth by grey spheres.}
    \label{fig:unitcell}
\end{figure}

The family of compounds RMIn$_5$ (M=Co, Rh, Ir), where R is a rare earth, presents a rich variety of electronic and magnetic properties ranging from heavy fermion behavior and anomalous superconductivity to complex magnetic states. These properties are closely related to the strong correlations on the R $4f$ electrons and to the quasi two-dimensionality of the Fermi surface.
These materials crystallize in a tetragonal structure that can be viewed as alternating MIn$_2$ and RIn$_3$ planes stacked along the $c$-axis (see Fig. \ref{fig:unitcell}), where the role of the transition metal M connecting the RIn$_3$ planes is central to determine the stability of the low temperature phase. 

The most puzzling features occur in the Ce-based compounds which present heavy fermion behavior at $T\lesssim$20 K. 
Correlation effects induce an enhancement of the electronic specific heat coefficient 
up to $1000$mJ/mol K$^2$ for CeCoIn$_5$ 
which is an ambient pressure superconductor below $T_C=2.3$K.\cite{0953-8984-13-17-103} CeIrIn$_5$ has its superconducting transition at $T_C=0.4$K while 
CeRhIn$_5$ is an antiferromagnet at ambient pressure with a N\'eel temperature $T_N=3.8$K.\cite{Movshovich2001} For $P>P_{cr}=1.77GPa$ the antiferromagnetic state of CeRhIn$_5$ is replaced by a superconducting state which coexists with magnetic order.\cite{Park2006} The less studied PuCoIn$_5$ and PuRhIn$_5$ compounds are heavy fermion superconductors and the highest superconducting temperature is also obtained for the Co-based compound with $T_C=2.5$K, while the reported value for PuRhIn$_5$ is $T_C=1.1$K.\cite{BauerPuCo,BauerUnpublished}

\begin{table*}[t]
    \begin{tabularx}{\textwidth}{@{}l *{10}{>{\centering\arraybackslash}X}@{}}
        \hline
\hline
 M\textbackslash R& Ce          & Nd         & Sm         & Gd       & Tb       & Dy       & Ho       & Er         & Tm      & Pu\\
\hline
    Co& ${\color{blue} \mathbf{2.3}}^a$ & $8^b$      & $11.9^c$   & $30^d$   & $30.2^e$ & $20^e$   & $10.5^e$ & $<2^e$   & $2.6^f$ & ${\color{blue}\mathbf{2.5}}^g$ \\
    Rh& $3.8^h$          & $11.6^i$   & $15^j$     & $39.9^k$ & $45.5^l$ & $28.1^m$ & $15.8^m$ & $3$--$4^k$ & $3.6^k$ & ${\color{blue}\mathbf{1.1}}^n$  \\
    Ir& ${\color{blue}\mathbf{0.4}}^r$ & $13.7^{i}$ & $14.3^{j}$ & $42^{j}$ & $41.4^o$ & --    & --      & --        & --     & --\\
\hline
\hline
\end{tabularx}
\caption{(Color online) N\'eel and superconducting transition temperatures for RMIn$_5$ compounds at ambient pressure. All temperatures are in K and bold numbers correspond to the superconducting transition temperature. Superscript letters correspond to References: $a=$ [\onlinecite{0953-8984-13-17-103}], $b=$[\onlinecite{Hudis2006}], $c$=[\onlinecite{JPSJS.76SA.62}], $d$=[\onlinecite{Betancourth2015}], $e$=[\onlinecite{Isikawa2004}], $f$=[\onlinecite{Huy20092425}], $g$=[\onlinecite{BauerPuCo}], $h$=[\onlinecite{PhysRevLett.84.4986}], $i$=[\onlinecite{PhysRevB.62.12266}], $j$=[\onlinecite{Pagliuso2001}], $k$=[\onlinecite{VanHieu2007}], $l$=[\onlinecite{PhysRevB.74.214404}], $m$=[\onlinecite{JPSJ.75.074708}], $n$=[\onlinecite{BauerUnpublished}], $p$=[\onlinecite{0295-5075-53-3-354}], $q$=[\onlinecite{serrano2006determinaccao}], and $r$=[\onlinecite{Movshovich2001}].}
\label{tab:tns}
\end{table*}
The metallic RMIn$_5$ (R=Nd, Sm, Gd, Tb, Dy, Ho, Er, and Tm, M=Co, Rh, Ir) compounds which order antiferromagnetically, show an interesting pattern in their N\'eel temperatures (see Table \ref{tab:tns}). 
For a given rare earth, the Rh and Ir based compounds have similar N\'eel temperatures while those based on Co order at a temperature $30\%-50\%$ lower.
In this article we address the above mentioned regularity and, to that aim, we focus our analysis on the magnetic behavior of the R=Gd compounds. The Gd-115 compounds are particularly appealing to study the role of the transition metal $d$ electrons on the magnetism, because of their relative simplicity. In these compounds the Gd$^{3+}$ ions are expected to be in a $S=7/2$, $L=0$ multiplet, and the crystal-field splitting effects are therefore expected to be much smaller than in Ce and other $L\neq 0$ rare earth analogues. Moreover, these materials do not show heavy fermion behavior which further simplifies the analysis. 
A deeper understanding of the behavior of the magnetic 115 compounds when the transition metal M is replaced may help asses the stability of the superconducting state in the Ce-115 and Pu-115 compounds. The superconductivity in these materials seems to be deeply associated with the magnetic properties and the highest superconducting temperatures are obtained for the Co-based compounds.

Total-energy calculations of the GdMIn$_5$ compounds, based on Density-functional-theory (DFT), indicate a ground state with magnetic moments localized at the Gd$^{3+}$ ions and allowed us to estimate the strength of the Gd-Gd magnetic interactions. We solved the resulting magnetic model to obtain the magnetic contribution to the specific heat, the magnetic susceptibility, and the N\'eel and Curie-Weiss temperatures. The excellent agreement obtained with the available experimental data validates our model and the calculated magnetic interaction parameters. 
As we show below, the fact that the GdCoIn$_5$ compound has a lower transition temperature than its Rh and Ir counterparts, can be associated with its strongly suppressed magnetic coupling between Gd$^{3+}$ ions located at different GdIn$_3$ planes. 

The reduced value for the interplane exchange coupling obtained in GdCoIn$_5$ is mainly due to a suppression of the hybridization between the Co 3$d$ and the Gd$^{3+}$ 5$d$ orbitals that mediate the interplane RKKY interaction between the Gd$^{3+}$ ion magnetic moments. A toy model considering a single effective $d$ orbital on the transition metal M and the Gd$^{3+}$ ions is able to qualitatively explain the behaviour of the interplane exchange coupling. The parameters for the model were calculated from a Wannier orbital analysis\cite{Anisimov2005} and roughly estimated from the average energy and the total width of the Co an Gd bands with dominating 3$d$ and 5$d$ character, respectively. The results from the two approaches lead to the same conclusions.   

The rest of the paper is organized as follows: 
In Sec. \ref{sec:magprop} we determine the magnetic structure and the coupling constants of the magnetic Hamiltonian for the three GdMIn$_5$ compounds and solve the model in the mean-field approximation and numerically using Quantum Monte Carlo (QMC). In Sec. \ref{sec:concl} we summarize our main results and conclusions.

\section{Magnetic properties} \label{sec:magprop}
In this Section we analyze the magnetic structure of the GdMIn$_5$ compounds. We propose a simple Hamiltonian to describe their magnetic properties and determine the model parameters through DFT calculations. 
To describe the temperature dependence of the magnetic properties we first treat the magnetic Hamiltonian in the mean-field approximation which allows a simple interpretation of the experimental data. We then include quantum fluctuations in a simplified model to obtain a quantitative description of the low temperature ($T\lesssim T_N$) experimental data.
\subsection{Technical details of the DFT calculations}
The total-energy calculations were performed using the generalized gradient approximation (GGA) of Perdew, Burke and Ernzerhof for the exchange and correlation functional as implemented in the Wien2K code.\cite{wien2k,Perdew1996}
A local Coulomb repulsion was included in the Gd $4f$ shell and treated using GGA+U which is a reasonable approximation for these highly localized states. GGA+U has been also used in previous calculations of compounds of the RMIn$_5$ family. \cite{Piekarz2005,PhysRevLett.96.237003,Zhu2012} 
Due to the localized character of the $4f$ electrons, the fully localized limit was used for the double counting correction.\cite{Anisimov1993} 
We described the local Coulomb and exchange interactions with a single effective local repulsion $U_{eff} = U - J_H = 6 eV$ as in bulk Gd. \cite{Yin2006,Petersen2006}
The APW+local orbitals method of the \textsc{WIEN2K} code was used for the basis function.\cite{wien2k} 1200 k-points were used in the irreducible Brioullin zone for the full optimization of the crystal structures, and 440 k-points for the $2\times2\times2$ supercell total-energy calculations of the different magnetics configurations.

\subsection{Magnetic structure of the ground state and coupling constants}

\begin{figure}[h]
  \centering
    \includegraphics[width=0.4\textwidth]{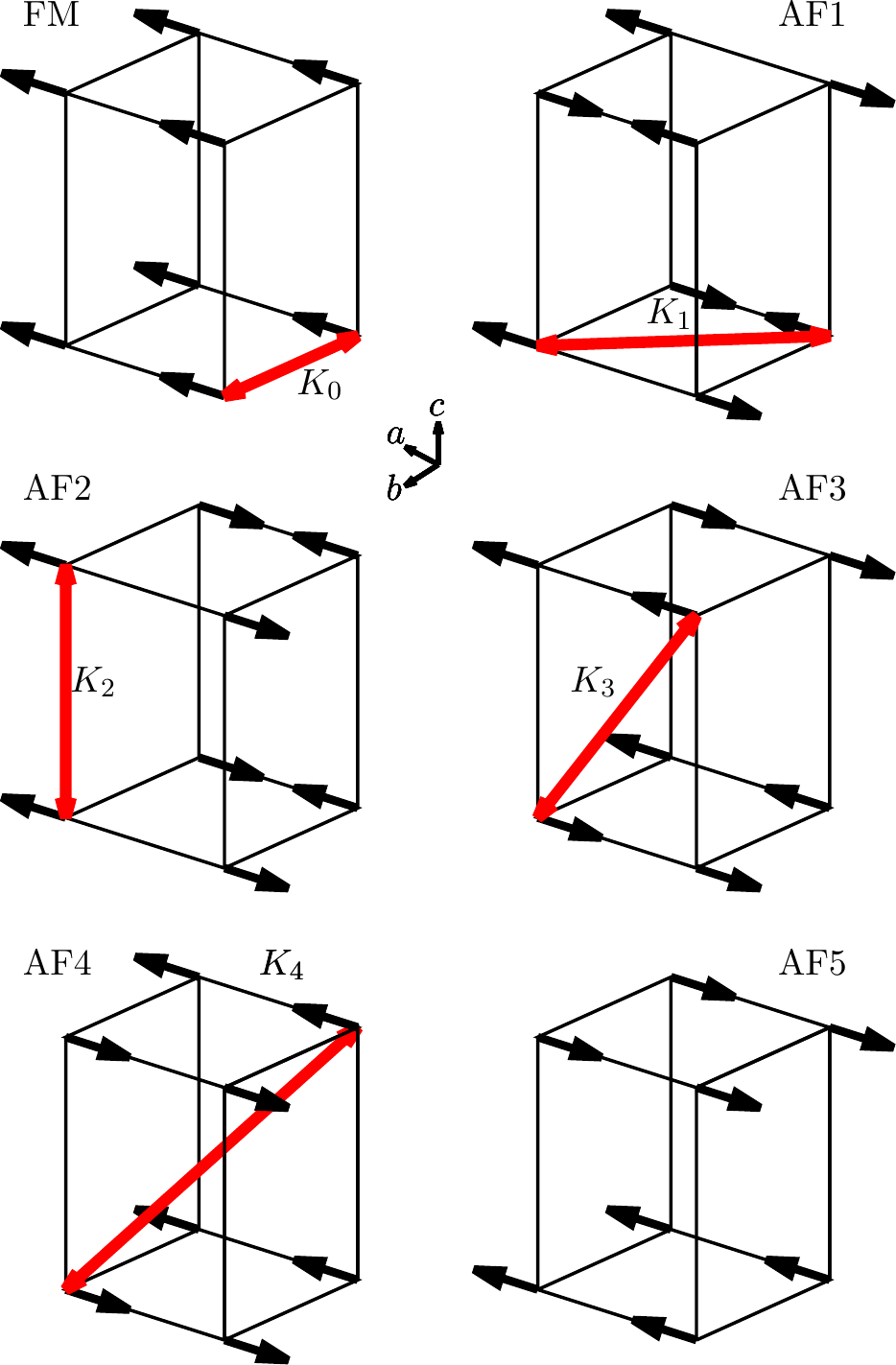}
  \caption{(Color online) Magnetic configurations proposed to determine the ground state and obtain the exchange coupling parameters. The Gd atoms are located at the vertices of the rectangular prism and the orientation of their magnetic moments is indicated by black arrows. The red arrows connect a pair of Gd atoms that are magnetically coupled through the exchange coupling parameters $K_0$, $K_1$, $K_2$, $K_3$, and $K_4$, as indicated in the figure. }
  \label{fig:configurations}
\end{figure}

We explored different static magnetic configurations for the magnetic moments which are presented in Fig. \ref{fig:configurations}.  
The lowest energy configuration for the three Gd compounds is antiferromagnetic (AF3) which corresponds to the measured structure in GdRhIn$_5$ via  resonant x-ray diffraction experiments, NdRhIn$_5$ in neutron diffraction experiments and the inferred structure of DyRhIn$_5$ and HoRhIn$_5$ in magnetization experiments.\cite{Granado2006,Chang2002,vanHieu1} 
This magnetic configuration is associated with the competition of the first-neighbour $K_0$ and the second neighbour $K_1$ antiferromagnetic exchange couplings that lead to ferromagnetic chains in the GdIn$_3$ plane and an antiferromagnetic interplane coupling $K_2$ that leads to an antiferromagnetic configuration between GdIn$_3$ planes.\cite{Granado2006,Chang2002,vanHieu1} The total energy for each magnetic configuration is presented in Table \ref{tab:totnrg}.

We assume that the magnetic interaction between the magnetic moments on the Gd$^{3+}$ ions can be described with the following Hamiltonian:
\begin{equation}
    \mathcal{H}= \sum_{i\neq j} K_{ij} \mathcal{J}_i\cdot \mathcal{J}_j
    \label{eq:magham}
\end{equation}
where $K_{ij}$ is the exchange coupling between the magnetic moments $\mathcal{J}_i$ and $\mathcal{J}_j$, and depends on the intraplane and interplane distances between Gd atoms. As it is indicated in Fig. \ref{fig:configurations}, $K_{ij}$ is equal to $K_0$ for nearest neighbors and $K_1$ for next-nearest-neighbors inside the GdIn$_3$ plane, and correspondingly to $K_2$, $K_3$ and $K_4$ for the interplane couplings. The dominating Gd-Gd magnetic exchange interactions are due to a RKKY coupling between the Gd's magnetic moments through exchange coupling with the conduction electrons. 

\begin{table}
	\centering
    \begin{tabularx}{\columnwidth}{@{}l *4{>{\centering\arraybackslash}X}@{}}
		\hline
		\hline
        & GdCoIn$_5$ &GdRhIn$_5$ &GdIrIn$_5$ \\
		\hline
		FM&	126&  145& 149 \\
		AF1&	62&   65& 56 \\
		AF2&	59&   95& 74 \\
		AF3&	0&    0&  0 \\
		AF4&	23&   50& 44 \\
		AF5&	125&  133& 128\\
		\hline
		\hline
	\end{tabularx}
    \caption{Relative energy $\Delta E$ (in K) with respect to the ground state for the magnetic configurations of Fig. \ref{fig:configurations}}
	\label{tab:totnrg}
\end{table}

In the absence of an applied magnetic field, the contribution per Gd atom to the total energy due to the magnetic interactions described in Eq. (\ref{eq:magham}) for the different configurations of Fig. \ref{fig:configurations} is given by:
\begin{equation}
    \begin{array}{r@{}l}
        E^m_{FM}/J^2 &={} 2 K_0  + 2 K_1 + K_2 + 4 K_3 + 4 K_4 \\
        E^m_{AF1} /J^2&={} -2 K_0  + 2 K_1 - K_2 + 4 K_3 - 4 K_4 \\
        E^m_{AF2}/J^2 &={} -2 K_0  + 2 K_1 + K_2 - 4 K_3 + 4 K_4 \\
        E^m_{AF3}/J^2 &={} -2 K_1  - K_2 + 4 K_4\\
        E^m_{AF4}/J^2 &={} -2 K_1  + K_2 - 4 K_4\\
        E^m_{AF5}/J^2 &={} 2 K_0  + 2 K_1 - K_2 - 4 K_3 - 4 K_4 
\end{array}
	\label{eq:couplings}
\end{equation}
where $J=7/2$ is the angular momentum of the Gd$^{3+}$ ion $4f$ electrons. The energy differences between magnetic configurations calculated from first principles can be combined with Eqs. (\ref{eq:couplings}) to obtain the coupling parameters $K_i$ solving a system of 5 linear equations. The results for the $K_i$ are presented in Table \ref{tab:exchcoup} and show some remarkable features. 
On the one hand, the interplane coupling $K_2$ is a factor $\sim 3$ smaller in GdCoIn$_5$ than in GdRhIn$_5$ and GdIrIn$_5$, while the other sizable couplings do not change significantly. 
On the other hand, $K_3$ and $K_4$ are much smaller than $K_2$ so that $K_2$ dominates the interplane coupling in the Rh an Ir compounds. This implies a more two-dimensional behavior of the magnetism in GdCoIn$_5$ than in GdRhIn$_5$ and GdIrIn$_5$ and, as we will see below, explains the lower N\'eel temperature observed in the Co-based compound. 
\begin{table}
    \centering
    %\begin{tabular}{|c|c|c|c|}
    \begin{tabularx}{\columnwidth}{@{}l *5{>{\centering\arraybackslash}X}@{}}
        \hline
        \hline
        & GdCoIn$_5$ & GdRhIn$_5$ & GdIrIn$_5$\\
        \hline
        $K_0$& $1.28\,$&$1.21\,$& $1.51$ \\
        $K_1$&  $1.64\,$&$1.74\,$& $1.63$\\
				$\bm{K_2}$&  $\mathbf{0.49}\,$&$\mathbf{1.43}\,$& $\mathbf{1.30}$ \\
        $K_3$&  $0.04\,$&$-0.01\,$& $0.02$ \\
        $K_4$&  $-0.11\,$&$-0.15\,$& $-0.12$ \\
        \hline
        $\theta$ & $63.3$ & $66.5$ &$75.2$ \\
        $\theta^{exp}$ & $\sim 50$ & $69$$^a$, $63.8$$^b$ &$64$$^a$ \\
        \hline
        $T_{N}^{MF}$ & $44.4$ & $57.6$ &$52.9$ \\
        $T_{N}^{QMC}$ & $32.3$ & $41.9$ &$38.4$ \\
        $T_{N}^{exp}$ & $30$ & $39$ &$40$ \\
        \hline
        \hline
    \end{tabularx}
    \caption{Calculated exchange couplings (in K) and the associated mean-field Curie-Weiss $\theta$ and N\'eel $T_N^{MF}$ temperatures. $T_N^{QMC}$ N\'eel temperature calculated using QMC on an effective model (see text). The experimental N\'eel $T_N^{exp}$ and Curie-Weiss $\theta^{exp}$ temperatures are presented as a reference. The superscripts indicate the references from which the experimental values were extracted: $a=$ [\onlinecite{Pagliuso2001}], $b=$[\onlinecite{VanHieu2007}].}
    \label{tab:exchcoup}
\end{table}
\subsection{Toy model for the interplane coupling $K_2$}
\begin{figure}[t]
	\begin{center}
        \includegraphics[width=0.45\textwidth]{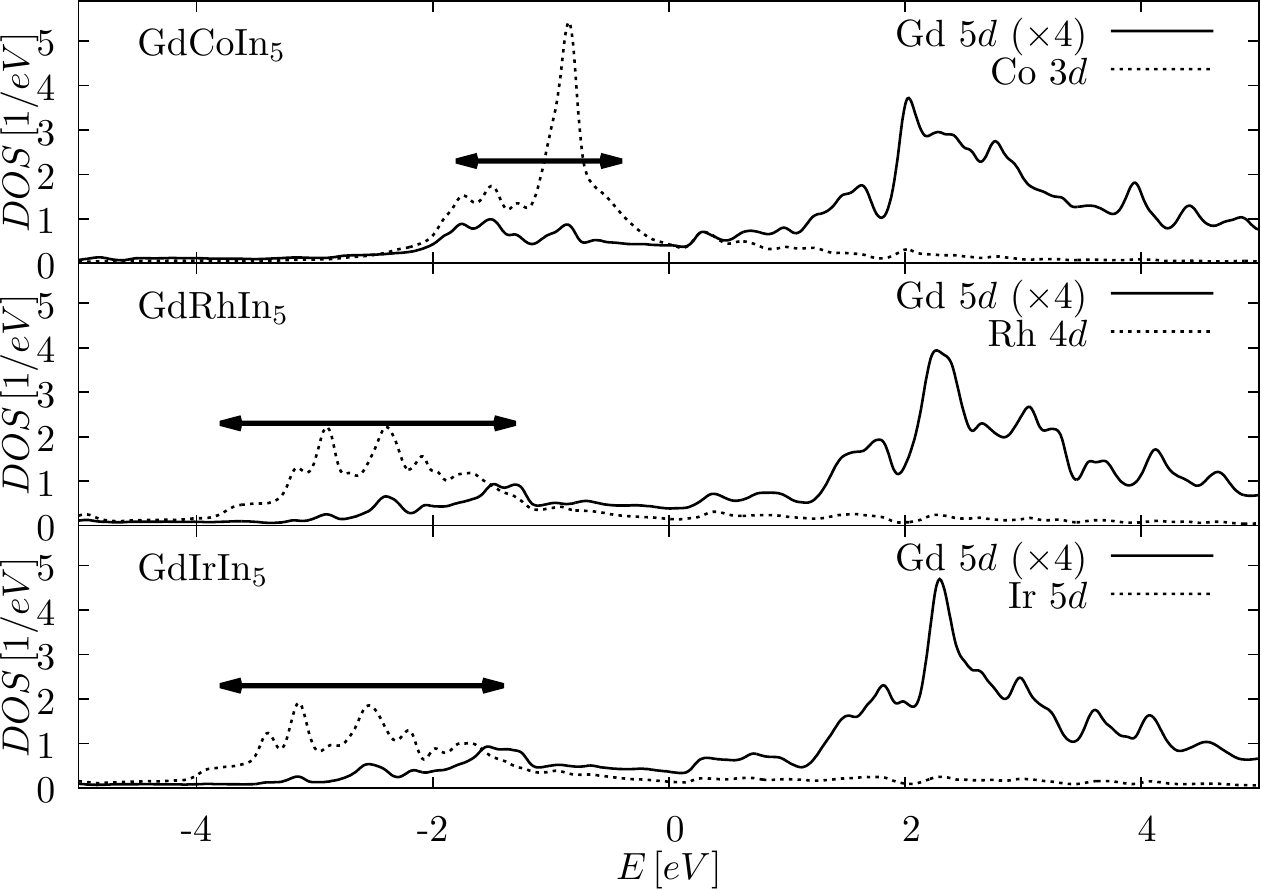}
    \end{center}
    \caption{Partial densities of states of the Gd$^{3+}$ $5d$ and the transition metal $d$ orbital for the three GdMIn$_5$ (M=Co, Rh, Ir) compounds. A small hybridization between the transition metal M and the Gd electrons can be deduced from the presence of Gd $d$ states at energies where the M $d$ electrons have a sizable DOS. The arrows indicate a rough estimation of the total bandwidth of the bands with mostly M $d$ character. The Gd$^{3+}$ $5d$ partial DOS has been multiplied by $4$. }
    \label{fig:partialDOS}
\end{figure}

The exchange couplings calculated in the previous Section stem from a RKKY mechanism mediated by the conduction electrons.\cite{Jensen&Mackintosh} The  Gd$^{3+}$ $4f$ electrons couple with the Gd$^{3+}$ $5d$ conduction electrons with a magnetic exchange coupling $J_{fd}$ that for a related material has been estimated to be $J_{fd}\sim 75meV$ (see Ref. [\onlinecite{Cabrera-Baez2014}]).
The almost empty Gd$^{3+}$ $5d$ orbitals have a small hybridization with the partially occupied transition metal $d$ orbitals (see Fig. \ref{fig:partialDOS}).
With these ingredients we construct a toy model, to describe the behavior of the RKKY coupling $K_2$, with parameters that could be estimated from experimental data. We consider a single effective level to represent the transition metal $d$ orbitals and another for the Gd$^{3+}$ $5d$ orbitals. While the In $5p$ orbitals contribute to the conduction electron bands, their inclusion in the toy model does not change qualitatively the results and will not be considered here.
To calculate the exchange coupling $K_2$ we consider two Gd atoms coupled via a single transition metal atom and calculate the energy of the parallel ($E_P$) and antiparallel ($E_{AP}$) configurations for the Gd$^{3+}$ $4f$ magnetic moments. The coupling is estimated as $K_2 \sim (E_{AP}-E_{P})/2J^2$.
The model hamiltonian is
\begin{equation}
    H_2 = \sum_\sigma E_{d\sigma}d_{\sigma}^\dagger d_{\sigma} + E_{c}\sum_\sigma c_{\sigma}^\dagger c_{\sigma} + t\sum_\sigma (d_\sigma^\dagger c_\sigma + h.c.)
    \label{toymodel}
\end{equation}
where $d_{\sigma}^\dagger$ ($c_{\sigma}^\dagger$) creates an electron with spin projection $\sigma=\pm$ along the z-axis on the Gd$^{3+}$ $5d$ (M $d$) effective orbital. The exchange coupling between the $4f$ and $5d$ Gd$^{3+}$ electrons is taken as a static field on the Gd$^{3+}$ $5d$ effective orbital making its energy spin dependent: $E_{d\sigma}=E_{d}\pm \sigma \Delta$ where the $+$ ($-$) sign corresponds to the Gd$^{3+}$ $4f$ magnetic moment being parallel (antiparallel) to the z-axis, and $\Delta=J\,J_{fd}$.
The model can be readily diagonalized and to lowest order in $t$ and $\Delta$:
\begin{equation}
	K_2\sim \frac{2\Delta^2 t^4}{J^2 (E_c-E_d)^5}
	\label{K2}
\end{equation}
where we have assumed that $t$ is a small parameter. 
The parameters $E_d$ and $E_c$ can be roughly estimated from the central weights of the bands with the highest Gd$^{3+}$ $5d$ and M $d$ character, respectively.  To obtain the ratio of hybridization parameters $t$ for two given compounds, we assumed it to be proportional to the ratio of the total bandwidths of the M $d$ bands in the corresponding compounds (see Fig. \ref{fig:partialDOS} for the estimation of the width of the M $d$ bands). The main assumption here is that the intraplane and interplane hybridizations change in the same proportion when the transition metal is changed. The estimation of the parameters from experimental data requires the measurement of the M $d$ and Gd $f$ partial DOS that could be obtained from resonant photoemission spectroscopy experiments.\cite{Levy2012}

The parameter $E_d \sim 3eV$ is nearly the same for the three compounds, while the level energy of the Co $3d$ orbital ($E_{c}^{Co}\sim -1.1eV$) is higher than the corresponding to Rh 4$d$ ($E_c^{Rh}\sim -2.5 eV$) and Ir 5$d$ ($E_c^{Ir}\sim -2.6 eV$). The hybridization $t$ is estimated (see Fig. \ref{fig:partialDOS}) to be a $\sim 40\%$ smaller in GdCoIn$_5$ than in the Rh and Ir compounds.
The model reproduces approximately the value of the ratios between the couplings $K_2$ of the three compounds. The reduced value of the $K_2$ exchange couplings in the Co-based compound is associated with a reduced hybridization $t$ compared to the Rh and Ir compounds which is partially compensated for the larger value of $E_c$ in GdCoIn$_5$.  
The coupling $K_2$ is expected to have similar values for GdRhIn$_5$ and GdIrIn$_5$ as the two materials have similar effective parameters and hybridization $t$.
Similar results are obtained estimating the parameters from a Wannier orbital analysis projecting the Hamiltonian on the partially occupied Gd$^{3+}$ $5d$ and M $d$ bands.\cite{Anisimov2005} 

\subsection{Finite temperatures} \label{sec:MF}
\begin{figure}[tb]
    \centering
\includegraphics[width=0.45\textwidth]{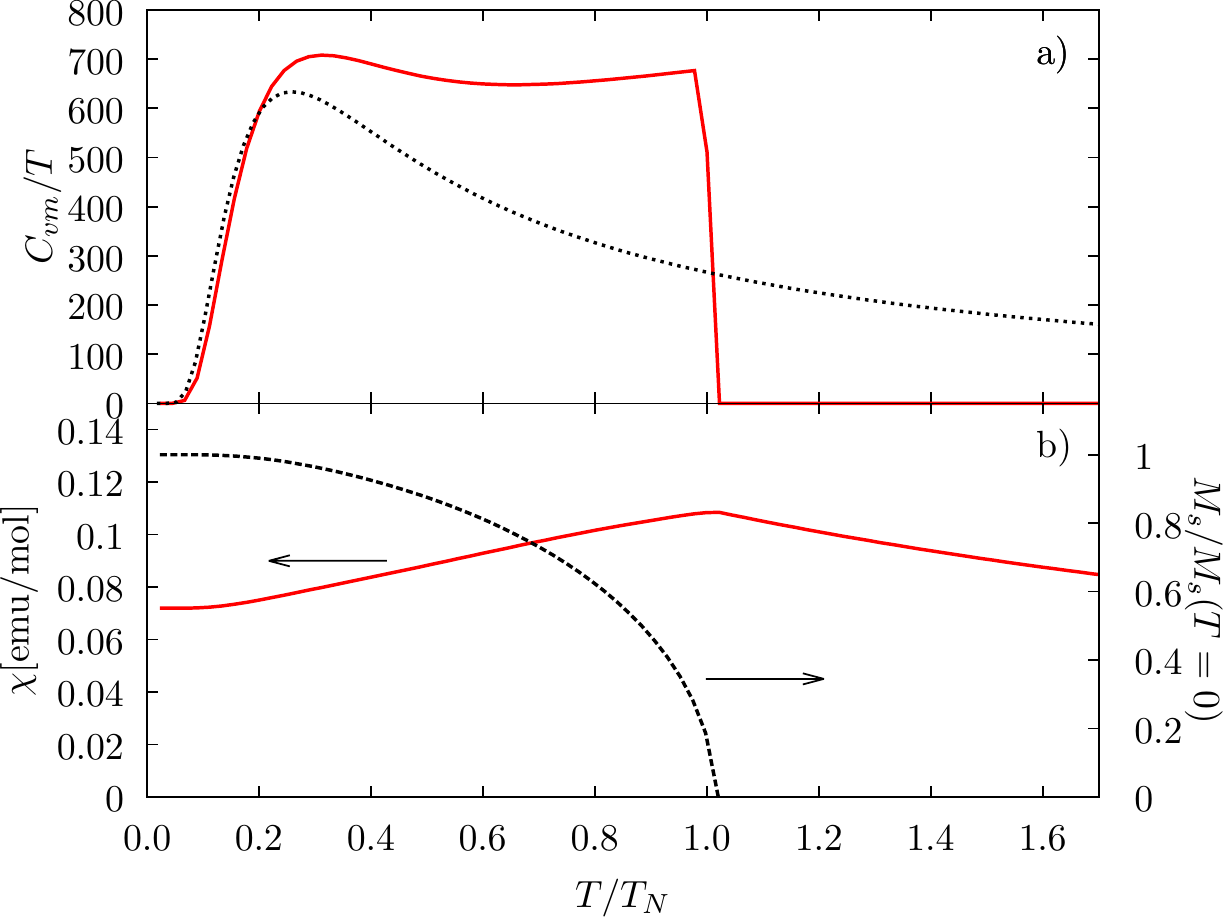}
\caption{(Color online) Mean-field solution of the magnetic Hamiltonian of Eq. \ref{eq:magham} using the calculated parameters for GdCoIn$_5$ from Table \ref{tab:exchcoup}. a) Specific heat $C_{vm}/T$ as a function of the temperature calculated using the coupling parameters for GdCoIn$_5$ (solid line) and for a two level system with level splitting $2k_BT_N/3$ (dotted line). b) Crystal averaged magnetic susceptibility (left axis) and staggered magnetization (right axis).} 
    \label{fig:MF}
\end{figure}

In this Section we analize the validity of the model Hamiltonian given by Eq. (\ref{eq:magham}) and the calculated exchange coupling parameters (see Table \ref{tab:exchcoup}) to describe the magnetic degrees of freedom in the GdMIn$_5$ compounds.
We solve the magnetic Hamiltonian using different approximations to obtain the magnetic susceptibility and the magnetic contribution to the specific heat and compare with the experimental data in the literature. 
We solved the magnetic Hamiltonian given by Eq. (\ref{eq:magham}) in the mean field (MF) approximation considering $8$ independent Gd$^{3+}$ magnetic moments. The model presents a paramagnetic to $C$-type antiferromagnet transition as the temperature is lowered below $T_N^{MF}=\frac{J(J+1)}{3}\left(4 K_1 + 2 K_2 - 8 K_4  \right)$.
 Figure \ref{fig:MF}a) presents the magnetic contribution to the specific heat which shows a discontinuity at the transition temperature and vanishes above $T_N$. 
The obtained behavior of $C_{vm}$ for temperatures above $T_N$ is a well known artifact of the mean-field solution.
The shoulder in $C_{vm}/T$ below the transition temperature is due to an increase in the staggered magnetization and the associated internal field as the temperature is lowered. 
The coupling to the internal field splits the different projections of each magnetic moment along the internal field axis and the higher energy projections are exponentially suppressed as the temperature is lowered. The peak in $C_{vm}/T$ at $T\sim 0.3 T_N$ can be associated with a Schotkky-like anomaly as $k_B T$ becomes of the order of the energy splitting $\Delta(T)$ between the two lowest laying states. This is illustrated by the specific heat contribution for a two level system with temperature independent energy splitting $\Delta(T=0)=\frac{2}{3} k_B T_N$ shown in Fig \ref{fig:MF}a).\cite{Blanco1991}     
The second order transition at $T_N$ is accompanied by the onset of the staggered magnetization $M_s$ [see Fig. \ref{fig:MF}b)] associated with an internal field $\mu_\text{B} H_{int} =\frac{3k_\text{B} T_N^{MF}}{J(J+1)} \langle \mathcal{J}\rangle$. As $T\to 0$ the staggered magnetization saturates and the specific heat is exponentially reduced at the mean-field level.  

\begin{figure}[t]
	\begin{center}
		\includegraphics[width=0.5\textwidth]{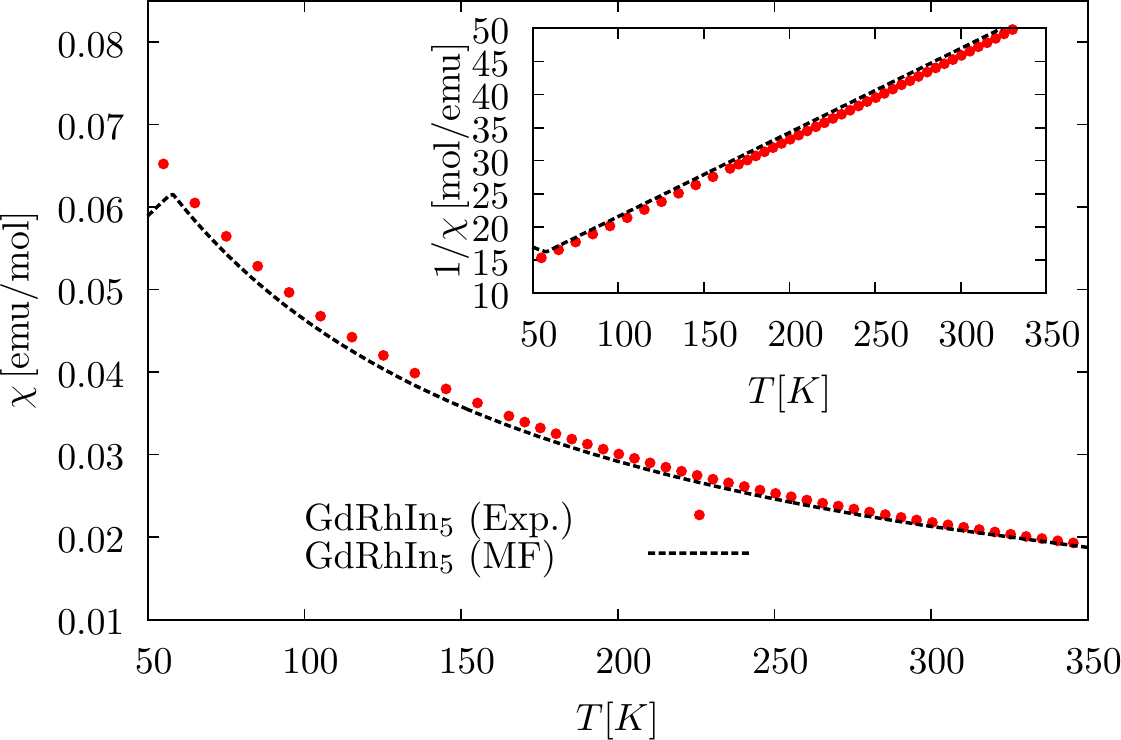}
	\end{center}
	\caption{(Color online) Crystal averaged magnetic susceptibility $\chi=M/B$, $B=0.1T$. Mean field and experimental\cite{Pagliuso2001} results for GdRhIn$_5$. Inset: inverse magnetic susceptibility as a function of the temperatures showing a Curie-Weiss behavior for $T>T_N$.  The mean-field results provide an accurate description at high temperatures.  }
	\label{fig:chi}
\end{figure}

The magnetic susceptibility is presented in Fig. \ref{fig:MF}b). At temperatures $T>T_N$, $\chi$ has a Curie-Weiss  behavior and decreases with decreasing temperature for $T<T_N$: $\chi = (g_\text{J} \mu_{\text{B}})^2 J(J+1)/3(T+\theta)$ where $\theta=\frac{J(J+1)}{3}\left(4 K_0 + 4 K_1 + 2 K_2 + 8 K_3 + 8 K_4  \right)$. The values of $\theta$ for the GdMIn$_5$ compounds using the calculated exchange couplings are presented in Table \ref{tab:exchcoup}. They present a good quantitative agreement with the experimental results.  
Figure \ref{fig:chi} presents a comparison between the mean-field and the experimental results for the magnetic susceptibility of GdRhIn$_5$ as a function of the temperature. There is an excellent agreement at temperatures above the N\'eel temperature where the material presents a Curie-Weiss behavior (see inset to Fig. \ref{fig:chi}). The N\'eel temperature is overestimated which, as we will see below, is a consequence of ignoring quantum fluctuations and the source of the low temperature disagreement between the mean-field magnetic susceptibility and the experimental data.

Although the mean-field solution is consistent with the experimental results it does not show some features in the specific heat like the power law behaviour of $C/T$ at small T (related with spin waves) nor the lambda divergence at the transition, and overestimates the transition temperature. To improve the description of the physical properties including quantum fluctuations we resort to a simplified magnetic model. We consider $J=7/2$ magnetic moments on a cubic lattice interacting through a first-neighbour antiferromagnetic exchange coupling $K_{eff}=\frac{3k_\text{B} T_N^{MF}}{J(J+1)}/z$, where $z=6$ is the number of neighbours. At the mean-field level the simplified model reproduces the transition temperature and the specific heat of the full model in the complete temperature range. It does not reproduce, however, the Curie-Weiss temperature which, within the mean field approximation, is now equal to $-T_N$.
The effective model can be solved numerically using Quantum Monte Carlo simulations in a finite size cluster. We considered sizes ranging from $L=6$ up to $L=30$ in a cubic lattice of $L\times L\times L$ sites and used a finite size scaling analysis to extrapolate to $L\to\infty$. Thermalization and measurements where performed with a temperature dependent number of sweep steps, ranging between $10^5$ and $10^6$ steps. The Quantum Monte Carlo simulations were performed using the ALPS\cite{alps} library, in particular the ``loop'' algorithm which allows the inclusion of external magnetic fields.

\begin{figure}[t]
	\begin{center}
		\includegraphics[width=0.5\textwidth]{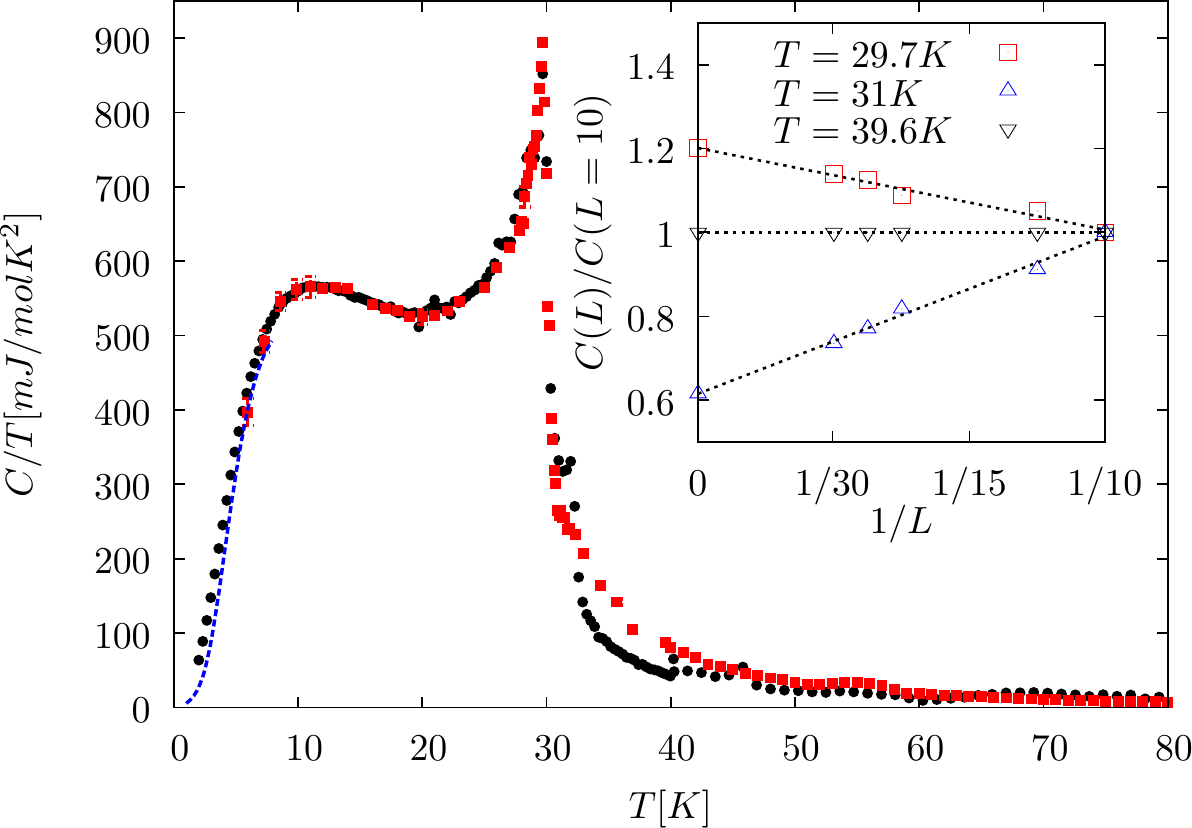}
	\end{center}
    \caption{(Color online) Magnetic contribution to the specific heat in GdCoIn$_5$. Experiment\cite{Betancourth2015} (black disks) and Quantum Monte Carlo (red squares). The experimental result was obtained substracting the calculated electron and phonon contributions to the specific heat. The blue dashed style line is the result of a spin-wave calculation valid at low temperatures.  
    Inset: finite size scaling of the Quantum Monte Carlo results for the specific heat. The lines are linear fits from which the $L \to \infty$ limit of the specific heat values were extracted. }
	\label{fig:QMC}
\end{figure}

The results for the magnetic contribution to the specific heat are presented in Fig. \ref{fig:QMC} together with the experimental data.\cite{Betancourth2015} The experimental $C_{vm}$ was obtained subtracting the theoretically obtained electronic and phonon contributions.\footnote{A detailed analysis of the phonon contribution to the specific heat in these compounds will be presented elsewhere.} The phonon contribution was corrected to account for anharmonic effects (see Refs. [\onlinecite{Betancourth2015}] and [\onlinecite{Wallace}]). We used a slightly reduced coupling $0.93\,K_{eff}$ in order to match the experimental transition temperature of GdCoIn$_5$. For a given coupling strength, the mean field solution overestimates de transition temperature by $\sim 50$\%.\cite{Oitmaa2004} The QMC results based on the DFT calculated magnetic interaction parameters predict a transition temperature within a $10$\% of the experimental observation for the three GdMIn$_5$ compounds. \cite{Oitmaa2004}
An excellent experiment-theory agreement is obtained in the full temperature range of $C_{vm}/T$, including the high temperature tail, the lambda transition, the plateau and the Schotkky-like anomaly. For temperatures close to $T_N$ the finite size effects are maximal and are corrected using a finite size scaling (see inset to Fig. \ref{fig:QMC}). 
At low temperatures ($T\lesssim 10$K) the error in the numerical calculations increases. To complete the description, we perform a spin-wave analysis.
The antiferromagnetic spin waves in a cubic lattice have a dispersion relation
\begin{equation}
	\omega(\mathbf{q}) = 2 K_{eff} J \sqrt{9-(\cos q_x+\cos q_y+\cos q_z)^2},
\end{equation}
and the specific heat is 
\begin{equation}
	C_{sw}(T)=R \frac{\partial}{\partial T}\int_0^\infty \omega(\mathbf{q}) n_b[\omega(\mathbf{q})]d^3q,
\end{equation}
where $n_b(\omega)$ is the Bose-Einstein distribution and $R$ is the universal gas constant. In the $T\to 0$ limit we have the expected temperature behavior $C_{sw}(T\to0)= R \frac{32 \pi^5}{15 (2 z)^{3/2}} \left(\frac{T}{J K_{eff}}\right)^3$. 
The resulting $C_{sw}$ allows us to extend the QMC results to low temperatures and is presented with a blue dashed style line in Fig. \ref{fig:QMC}.

\section{Summary and Conclusions}\label{sec:concl}
We analyzed the role of the transition metal in the RMIn$_5$ family of compounds. We focused our analysis on the GdMIn$_5$ (M=Co, Rh, Ir) which order magnetically at low temperatures. 
Based on DFT calculations we obtained the parameters of a magnetic Hamiltonian that we solved in the mean-field approximation and numerically through quantum Monte Carlo calculations. We obtained an excellent agreement with the experimental transition and the Curie-Weiss temperatures, as well as with the specific heat in the full temperature range. 
Our results show that the source of the diminished N\'eel temperature observed in the Co-based 115 compounds, compared to the Rh- or Ir-based compounds is associated with a reduced exchange coupling between the magnetic moments of the Gd$^{3+}$ ions located on different GdIn$_3$ planes. This reduced interplane RKKY coupling is associated with a reduced hybridization between the Gd $5d$ and the Co $3d$ electrons. 
We believe that the reduced coupling between planes may be the source of reduced $T_N$ on other compounds of the series. Our results indicate that the magnetism in the Co-based materials has a more two-dimensional character than in their Rh and Ir counterparts. 
This behavior may also help understand the larger superconducting transition temperatures observed in the Co-based Ce-115 and Pu-115 compounds, compared with the Rh- or Ir-based counterparts.

Future work includes extending our theoretical and experimental analysis to the Tb-115 compounds that present sizable crystal-field effects.

\acknowledgements
We thank P. Pagliuso and R. Lora Serrano for providing experimental data for the GdCoIn$_5$ and GdRhIn$_5$ compounds. We acknowledge financial support from CONICET grants PIP00832, PIP00273, and PIP0702, ANPCyT grant PICT07-00812, SeCTyP-UNCuyo grant 06/C393.
\bibliography{phon,115b,devictor}{}

\begin{thebibliography}{38}%
\makeatletter
\providecommand \@ifxundefined [1]{%
 \@ifx{#1\undefined}
}%
\providecommand \@ifnum [1]{%
 \ifnum #1\expandafter \@firstoftwo
 \else \expandafter \@secondoftwo
 \fi
}%
\providecommand \@ifx [1]{%
 \ifx #1\expandafter \@firstoftwo
 \else \expandafter \@secondoftwo
 \fi
}%
\providecommand \natexlab [1]{#1}%
\providecommand \enquote  [1]{``#1''}%
\providecommand \bibnamefont  [1]{#1}%
\providecommand \bibfnamefont [1]{#1}%
\providecommand \citenamefont [1]{#1}%
\providecommand \href@noop [0]{\@secondoftwo}%
\providecommand \href [0]{\begingroup \@sanitize@url \@href}%
\providecommand \@href[1]{\@@startlink{#1}\@@href}%
\providecommand \@@href[1]{\endgroup#1\@@endlink}%
\providecommand \@sanitize@url [0]{\catcode `\\12\catcode `\$12\catcode
  `\&12\catcode `\#12\catcode `\^12\catcode `\_12\catcode `\%12\relax}%
\providecommand \@@startlink[1]{}%
\providecommand \@@endlink[0]{}%
\providecommand \url  [0]{\begingroup\@sanitize@url \@url }%
\providecommand \@url [1]{\endgroup\@href {#1}{\urlprefix }}%
\providecommand \urlprefix  [0]{URL }%
\providecommand \Eprint [0]{\href }%
\providecommand \doibase [0]{http://dx.doi.org/}%
\providecommand \selectlanguage [0]{\@gobble}%
\providecommand \bibinfo  [0]{\@secondoftwo}%
\providecommand \bibfield  [0]{\@secondoftwo}%
\providecommand \translation [1]{[#1]}%
\providecommand \BibitemOpen [0]{}%
\providecommand \bibitemStop [0]{}%
\providecommand \bibitemNoStop [0]{.\EOS\space}%
\providecommand \EOS [0]{\spacefactor3000\relax}%
\providecommand \BibitemShut  [1]{\csname bibitem#1\endcsname}%
\let\auto@bib@innerbib\@empty
%</preamble>
\bibitem [{\citenamefont {Petrovic}\ \emph
  {et~al.}(2001{\natexlab{a}})\citenamefont {Petrovic}, \citenamefont
  {Pagliuso}, \citenamefont {Hundley}, \citenamefont {Movshovich},
  \citenamefont {Sarrao}, \citenamefont {Thompson}, \citenamefont {Fisk},\ and\
  \citenamefont {Monthoux}}]{0953-8984-13-17-103}%
  \BibitemOpen
  \bibfield  {author} {\bibinfo {author} {\bibfnamefont {C.}~\bibnamefont
  {Petrovic}}, \bibinfo {author} {\bibfnamefont {P.~G.}\ \bibnamefont
  {Pagliuso}}, \bibinfo {author} {\bibfnamefont {M.~F.}\ \bibnamefont
  {Hundley}}, \bibinfo {author} {\bibfnamefont {R.}~\bibnamefont {Movshovich}},
  \bibinfo {author} {\bibfnamefont {J.~L.}\ \bibnamefont {Sarrao}}, \bibinfo
  {author} {\bibfnamefont {J.~D.}\ \bibnamefont {Thompson}}, \bibinfo {author}
  {\bibfnamefont {Z.}~\bibnamefont {Fisk}}, \ and\ \bibinfo {author}
  {\bibfnamefont {P.}~\bibnamefont {Monthoux}},\ }\href
  {http://stacks.iop.org/0953-8984/13/i=17/a=103} {\bibfield  {journal}
  {\bibinfo  {journal} {J. Phys: Condens. Matter}\ }\textbf {\bibinfo {volume}
  {13}},\ \bibinfo {pages} {L337} (\bibinfo {year}
  {2001}{\natexlab{a}})}\BibitemShut {NoStop}%
\bibitem [{\citenamefont {Movshovich}\ \emph {et~al.}(2001)\citenamefont
  {Movshovich}, \citenamefont {Jaime}, \citenamefont {Thompson}, \citenamefont
  {Petrovic}, \citenamefont {Fisk}, \citenamefont {Pagliuso},\ and\
  \citenamefont {Sarrao}}]{Movshovich2001}%
  \BibitemOpen
  \bibfield  {author} {\bibinfo {author} {\bibfnamefont {R.}~\bibnamefont
  {Movshovich}}, \bibinfo {author} {\bibfnamefont {M.}~\bibnamefont {Jaime}},
  \bibinfo {author} {\bibfnamefont {J.~D.}\ \bibnamefont {Thompson}}, \bibinfo
  {author} {\bibfnamefont {C.}~\bibnamefont {Petrovic}}, \bibinfo {author}
  {\bibfnamefont {Z.}~\bibnamefont {Fisk}}, \bibinfo {author} {\bibfnamefont
  {P.~G.}\ \bibnamefont {Pagliuso}}, \ and\ \bibinfo {author} {\bibfnamefont
  {J.~L.}\ \bibnamefont {Sarrao}},\ }\href {\doibase
  10.1103/PhysRevLett.86.5152} {\bibfield  {journal} {\bibinfo  {journal}
  {Phys. Rev. Lett.}\ }\textbf {\bibinfo {volume} {86}},\ \bibinfo {pages}
  {5152} (\bibinfo {year} {2001})}\BibitemShut {NoStop}%
\bibitem [{\citenamefont {Park}\ \emph {et~al.}(2006)\citenamefont {Park},
  \citenamefont {Ronning}, \citenamefont {Yuan}, \citenamefont {Salamon},
  \citenamefont {Movshovich}, \citenamefont {Sarrao},\ and\ \citenamefont
  {Thompson}}]{Park2006}%
  \BibitemOpen
  \bibfield  {author} {\bibinfo {author} {\bibfnamefont {T.}~\bibnamefont
  {Park}}, \bibinfo {author} {\bibfnamefont {F.}~\bibnamefont {Ronning}},
  \bibinfo {author} {\bibfnamefont {H.~Q.}\ \bibnamefont {Yuan}}, \bibinfo
  {author} {\bibfnamefont {M.~B.}\ \bibnamefont {Salamon}}, \bibinfo {author}
  {\bibfnamefont {R.}~\bibnamefont {Movshovich}}, \bibinfo {author}
  {\bibfnamefont {J.}~\bibnamefont {Sarrao}}, \ and\ \bibinfo {author}
  {\bibfnamefont {J.~D.}\ \bibnamefont {Thompson}},\ }\href
  {http://www.nature.com/nature/journal/v440/n7080/suppinfo/nature04571_S1.html}
  {\bibfield  {journal} {\bibinfo  {journal} {Nature}\ }\textbf {\bibinfo
  {volume} {440}},\ \bibinfo {pages} {65} (\bibinfo {year} {2006})}\BibitemShut
  {NoStop}%
\bibitem [{\citenamefont {Bauer}\ \emph {et~al.}(2012)\citenamefont {Bauer},
  \citenamefont {Altarawneh}, \citenamefont {Tobash}, \citenamefont {Gofryk},
  \citenamefont {Ayala-Valenzuela}, \citenamefont {Mitchell}, \citenamefont
  {McDonald}, \citenamefont {Mielke}, \citenamefont {Ronning}, \citenamefont
  {Griveau}, \citenamefont {Colineau}, \citenamefont {Eloirdi}, \citenamefont
  {Caciuffo}, \citenamefont {Scott}, \citenamefont {Janka}, \citenamefont
  {Kauzlarich},\ and\ \citenamefont {Thompson}}]{BauerPuCo}%
  \BibitemOpen
  \bibfield  {author} {\bibinfo {author} {\bibfnamefont {E.~D.}\ \bibnamefont
  {Bauer}}, \bibinfo {author} {\bibfnamefont {M.~M.}\ \bibnamefont
  {Altarawneh}}, \bibinfo {author} {\bibfnamefont {P.~H.}\ \bibnamefont
  {Tobash}}, \bibinfo {author} {\bibfnamefont {K.}~\bibnamefont {Gofryk}},
  \bibinfo {author} {\bibfnamefont {O.~E.}\ \bibnamefont {Ayala-Valenzuela}},
  \bibinfo {author} {\bibfnamefont {J.~N.}\ \bibnamefont {Mitchell}}, \bibinfo
  {author} {\bibfnamefont {R.~D.}\ \bibnamefont {McDonald}}, \bibinfo {author}
  {\bibfnamefont {C.~H.}\ \bibnamefont {Mielke}}, \bibinfo {author}
  {\bibfnamefont {F.}~\bibnamefont {Ronning}}, \bibinfo {author} {\bibfnamefont
  {J.-C.}\ \bibnamefont {Griveau}}, \bibinfo {author} {\bibfnamefont
  {E.}~\bibnamefont {Colineau}}, \bibinfo {author} {\bibfnamefont
  {R.}~\bibnamefont {Eloirdi}}, \bibinfo {author} {\bibfnamefont
  {R.}~\bibnamefont {Caciuffo}}, \bibinfo {author} {\bibfnamefont {B.~L.}\
  \bibnamefont {Scott}}, \bibinfo {author} {\bibfnamefont {O.}~\bibnamefont
  {Janka}}, \bibinfo {author} {\bibfnamefont {S.~M.}\ \bibnamefont
  {Kauzlarich}}, \ and\ \bibinfo {author} {\bibfnamefont {J.~D.}\ \bibnamefont
  {Thompson}},\ }\href {http://stacks.iop.org/0953-8984/24/i=5/a=052206}
  {\bibfield  {journal} {\bibinfo  {journal} {J. Phys: Condens. Matter}\
  }\textbf {\bibinfo {volume} {24}},\ \bibinfo {pages} {052206} (\bibinfo
  {year} {2012})}\BibitemShut {NoStop}%
\bibitem [{Bau()}]{BauerUnpublished}%
  \BibitemOpen
  \href@noop {} {\ }\bibinfo {note} {Bauer, E. $et. $ $al.$,
  unpublished.}\BibitemShut {Stop}%
\bibitem [{\citenamefont {Hudis}\ \emph {et~al.}(2006)\citenamefont {Hudis},
  \citenamefont {Hu}, \citenamefont {Broholm}, \citenamefont {Mitrovi\'{c}},\
  and\ \citenamefont {Petrovic}}]{Hudis2006}%
  \BibitemOpen
  \bibfield  {author} {\bibinfo {author} {\bibfnamefont {J.}~\bibnamefont
  {Hudis}}, \bibinfo {author} {\bibfnamefont {R.}~\bibnamefont {Hu}}, \bibinfo
  {author} {\bibfnamefont {C.}~\bibnamefont {Broholm}}, \bibinfo {author}
  {\bibfnamefont {V.}~\bibnamefont {Mitrovi\'{c}}}, \ and\ \bibinfo {author}
  {\bibfnamefont {C.}~\bibnamefont {Petrovic}},\ }\href
  {http://www.sciencedirect.com/science/article/pii/S0304885306007657}
  {\bibfield  {journal} {\bibinfo  {journal} {J. Magn. Magn. Mater.}\ }\textbf
  {\bibinfo {volume} {307}},\ \bibinfo {pages} {301} (\bibinfo {year}
  {2006})}\BibitemShut {NoStop}%
\bibitem [{\citenamefont {Koeda}\ \emph {et~al.}(2007)\citenamefont {Koeda},
  \citenamefont {Hedo}, \citenamefont {Fujiwara}, \citenamefont {Uwatoko},
  \citenamefont {Sadamasa},\ and\ \citenamefont {Inada}}]{JPSJS.76SA.62}%
  \BibitemOpen
  \bibfield  {author} {\bibinfo {author} {\bibfnamefont {M.}~\bibnamefont
  {Koeda}}, \bibinfo {author} {\bibfnamefont {M.}~\bibnamefont {Hedo}},
  \bibinfo {author} {\bibfnamefont {T.}~\bibnamefont {Fujiwara}}, \bibinfo
  {author} {\bibfnamefont {Y.}~\bibnamefont {Uwatoko}}, \bibinfo {author}
  {\bibfnamefont {T.}~\bibnamefont {Sadamasa}}, \ and\ \bibinfo {author}
  {\bibfnamefont {Y.}~\bibnamefont {Inada}},\ }\href {\doibase
  10.1143/JPSJS.76SA.62} {\bibfield  {journal} {\bibinfo  {journal} {Journal of
  the Physical Society of Japan}\ }\textbf {\bibinfo {volume} {76}},\ \bibinfo
  {pages} {62} (\bibinfo {year} {2007})},\ \Eprint
  {http://arxiv.org/abs/http://dx.doi.org/10.1143/JPSJS.76SA.62}
  {http://dx.doi.org/10.1143/JPSJS.76SA.62} \BibitemShut {NoStop}%
\bibitem [{\citenamefont {Betancourth}\ \emph {et~al.}(2015)\citenamefont
  {Betancourth}, \citenamefont {Facio}, \citenamefont {Pedrazzini},
  \citenamefont {Jesus}, \citenamefont {Pagliuso}, \citenamefont {Vildosola},
  \citenamefont {Cornaglia}, \citenamefont {García},\ and\ \citenamefont
  {Correa}}]{Betancourth2015}%
  \BibitemOpen
  \bibfield  {author} {\bibinfo {author} {\bibfnamefont {D.}~\bibnamefont
  {Betancourth}}, \bibinfo {author} {\bibfnamefont {J.}~\bibnamefont {Facio}},
  \bibinfo {author} {\bibfnamefont {P.}~\bibnamefont {Pedrazzini}}, \bibinfo
  {author} {\bibfnamefont {C.}~\bibnamefont {Jesus}}, \bibinfo {author}
  {\bibfnamefont {P.}~\bibnamefont {Pagliuso}}, \bibinfo {author}
  {\bibfnamefont {V.}~\bibnamefont {Vildosola}}, \bibinfo {author}
  {\bibfnamefont {P.~S.}\ \bibnamefont {Cornaglia}}, \bibinfo {author}
  {\bibfnamefont {D.}~\bibnamefont {García}}, \ and\ \bibinfo {author}
  {\bibfnamefont {V.}~\bibnamefont {Correa}},\ }\href {\doibase
  http://dx.doi.org/10.1016/j.jmmm.2014.09.024} {\bibfield  {journal} {\bibinfo
   {journal} {Journal of Magnetism and Magnetic Materials}\ }\textbf {\bibinfo
  {volume} {374}},\ \bibinfo {pages} {744 } (\bibinfo {year}
  {2015})}\BibitemShut {NoStop}%
\bibitem [{\citenamefont {Isikawa}\ \emph {et~al.}(2004)\citenamefont
  {Isikawa}, \citenamefont {Kato}, \citenamefont {Mitsuda}, \citenamefont
  {Mizushima},\ and\ \citenamefont {Kuwai}}]{Isikawa2004}%
  \BibitemOpen
  \bibfield  {author} {\bibinfo {author} {\bibfnamefont {Y.}~\bibnamefont
  {Isikawa}}, \bibinfo {author} {\bibfnamefont {D.}~\bibnamefont {Kato}},
  \bibinfo {author} {\bibfnamefont {A.}~\bibnamefont {Mitsuda}}, \bibinfo
  {author} {\bibfnamefont {T.}~\bibnamefont {Mizushima}}, \ and\ \bibinfo
  {author} {\bibfnamefont {T.}~\bibnamefont {Kuwai}},\ }\href {\doibase
  10.1016/j.jmmm.2003.12.1021} {\bibfield  {journal} {\bibinfo  {journal} {J.
  Magn. Magn. Mater.}\ }\textbf {\bibinfo {volume} {272-276}},\ \bibinfo
  {pages} {635} (\bibinfo {year} {2004})}\BibitemShut {NoStop}%
\bibitem [{\citenamefont {Huy}\ \emph {et~al.}(2009)\citenamefont {Huy},
  \citenamefont {Noguchi}, \citenamefont {Hieu}, \citenamefont {Shao},
  \citenamefont {Sugimoto},\ and\ \citenamefont {Ishida}}]{Huy20092425}%
  \BibitemOpen
  \bibfield  {author} {\bibinfo {author} {\bibfnamefont {H.~T.}\ \bibnamefont
  {Huy}}, \bibinfo {author} {\bibfnamefont {S.}~\bibnamefont {Noguchi}},
  \bibinfo {author} {\bibfnamefont {N.~V.}\ \bibnamefont {Hieu}}, \bibinfo
  {author} {\bibfnamefont {X.}~\bibnamefont {Shao}}, \bibinfo {author}
  {\bibfnamefont {T.}~\bibnamefont {Sugimoto}}, \ and\ \bibinfo {author}
  {\bibfnamefont {T.}~\bibnamefont {Ishida}},\ }\href {\doibase
  http://dx.doi.org/10.1016/j.jmmm.2009.02.123} {\bibfield  {journal} {\bibinfo
   {journal} {Journal of Magnetism and Magnetic Materials}\ }\textbf {\bibinfo
  {volume} {321}},\ \bibinfo {pages} {2425 } (\bibinfo {year}
  {2009})}\BibitemShut {NoStop}%
\bibitem [{\citenamefont {Hegger}\ \emph {et~al.}(2000)\citenamefont {Hegger},
  \citenamefont {Petrovic}, \citenamefont {Moshopoulou}, \citenamefont
  {Hundley}, \citenamefont {Sarrao}, \citenamefont {Fisk},\ and\ \citenamefont
  {Thompson}}]{PhysRevLett.84.4986}%
  \BibitemOpen
  \bibfield  {author} {\bibinfo {author} {\bibfnamefont {H.}~\bibnamefont
  {Hegger}}, \bibinfo {author} {\bibfnamefont {C.}~\bibnamefont {Petrovic}},
  \bibinfo {author} {\bibfnamefont {E.~G.}\ \bibnamefont {Moshopoulou}},
  \bibinfo {author} {\bibfnamefont {M.~F.}\ \bibnamefont {Hundley}}, \bibinfo
  {author} {\bibfnamefont {J.~L.}\ \bibnamefont {Sarrao}}, \bibinfo {author}
  {\bibfnamefont {Z.}~\bibnamefont {Fisk}}, \ and\ \bibinfo {author}
  {\bibfnamefont {J.~D.}\ \bibnamefont {Thompson}},\ }\href {\doibase
  10.1103/PhysRevLett.84.4986} {\bibfield  {journal} {\bibinfo  {journal}
  {Phys. Rev. Lett.}\ }\textbf {\bibinfo {volume} {84}},\ \bibinfo {pages}
  {4986} (\bibinfo {year} {2000})}\BibitemShut {NoStop}%
\bibitem [{\citenamefont {Pagliuso}\ \emph {et~al.}(2000)\citenamefont
  {Pagliuso}, \citenamefont {Thompson}, \citenamefont {Hundley},\ and\
  \citenamefont {Sarrao}}]{PhysRevB.62.12266}%
  \BibitemOpen
  \bibfield  {author} {\bibinfo {author} {\bibfnamefont {P.~G.}\ \bibnamefont
  {Pagliuso}}, \bibinfo {author} {\bibfnamefont {J.~D.}\ \bibnamefont
  {Thompson}}, \bibinfo {author} {\bibfnamefont {M.~F.}\ \bibnamefont
  {Hundley}}, \ and\ \bibinfo {author} {\bibfnamefont {J.~L.}\ \bibnamefont
  {Sarrao}},\ }\href {\doibase 10.1103/PhysRevB.62.12266} {\bibfield  {journal}
  {\bibinfo  {journal} {Phys. Rev. B}\ }\textbf {\bibinfo {volume} {62}},\
  \bibinfo {pages} {12266} (\bibinfo {year} {2000})}\BibitemShut {NoStop}%
\bibitem [{\citenamefont {Pagliuso}\ \emph {et~al.}(2001)\citenamefont
  {Pagliuso}, \citenamefont {Thompson}, \citenamefont {Hundley}, \citenamefont
  {Sarrao},\ and\ \citenamefont {Fisk}}]{Pagliuso2001}%
  \BibitemOpen
  \bibfield  {author} {\bibinfo {author} {\bibfnamefont {P.~G.}\ \bibnamefont
  {Pagliuso}}, \bibinfo {author} {\bibfnamefont {J.~D.}\ \bibnamefont
  {Thompson}}, \bibinfo {author} {\bibfnamefont {M.~F.}\ \bibnamefont
  {Hundley}}, \bibinfo {author} {\bibfnamefont {J.~L.}\ \bibnamefont {Sarrao}},
  \ and\ \bibinfo {author} {\bibfnamefont {Z.}~\bibnamefont {Fisk}},\ }\href
  {\doibase 10.1103/PhysRevB.63.054426} {\bibfield  {journal} {\bibinfo
  {journal} {Phys. Rev. B}\ }\textbf {\bibinfo {volume} {63}},\ \bibinfo
  {pages} {054426} (\bibinfo {year} {2001})}\BibitemShut {NoStop}%
\bibitem [{\citenamefont {{Van Hieu}}\ \emph {et~al.}(2007)\citenamefont {{Van
  Hieu}}, \citenamefont {Takeuchi}, \citenamefont {Shishido}, \citenamefont
  {Tonohiro}, \citenamefont {Yamada}, \citenamefont {Nakashima}, \citenamefont
  {Sugiyama}, \citenamefont {Settai}, \citenamefont {{D. Matsuda}},
  \citenamefont {Haga}, \citenamefont {Hagiwara}, \citenamefont {Kindo},
  \citenamefont {Araki}, \citenamefont {Nozue},\ and\ \citenamefont
  {Onuki}}]{VanHieu2007}%
  \BibitemOpen
  \bibfield  {author} {\bibinfo {author} {\bibfnamefont {N.}~\bibnamefont {{Van
  Hieu}}}, \bibinfo {author} {\bibfnamefont {T.}~\bibnamefont {Takeuchi}},
  \bibinfo {author} {\bibfnamefont {H.}~\bibnamefont {Shishido}}, \bibinfo
  {author} {\bibfnamefont {C.}~\bibnamefont {Tonohiro}}, \bibinfo {author}
  {\bibfnamefont {T.}~\bibnamefont {Yamada}}, \bibinfo {author} {\bibfnamefont
  {H.}~\bibnamefont {Nakashima}}, \bibinfo {author} {\bibfnamefont
  {K.}~\bibnamefont {Sugiyama}}, \bibinfo {author} {\bibfnamefont
  {R.}~\bibnamefont {Settai}}, \bibinfo {author} {\bibfnamefont
  {T.}~\bibnamefont {{D. Matsuda}}}, \bibinfo {author} {\bibfnamefont
  {Y.}~\bibnamefont {Haga}}, \bibinfo {author} {\bibfnamefont {M.}~\bibnamefont
  {Hagiwara}}, \bibinfo {author} {\bibfnamefont {K.}~\bibnamefont {Kindo}},
  \bibinfo {author} {\bibfnamefont {S.}~\bibnamefont {Araki}}, \bibinfo
  {author} {\bibfnamefont {Y.}~\bibnamefont {Nozue}}, \ and\ \bibinfo {author}
  {\bibfnamefont {Y.}~\bibnamefont {Onuki}},\ }\href {\doibase
  10.1143/JPSJ.76.064702} {\bibfield  {journal} {\bibinfo  {journal} {J. Phys.
  Soc. Japan}\ }\textbf {\bibinfo {volume} {76}},\ \bibinfo {pages} {064702}
  (\bibinfo {year} {2007})}\BibitemShut {NoStop}%
\bibitem [{\citenamefont {Lora-Serrano}\ \emph {et~al.}(2006)\citenamefont
  {Lora-Serrano}, \citenamefont {Giles}, \citenamefont {Granado}, \citenamefont
  {Garcia}, \citenamefont {Miranda}, \citenamefont {Ag\"uero}, \citenamefont
  {Mendon\ifmmode \mbox{\c{c}}\else~\c{c}\fi{}a Ferreira}, \citenamefont
  {Duque},\ and\ \citenamefont {Pagliuso}}]{PhysRevB.74.214404}%
  \BibitemOpen
  \bibfield  {author} {\bibinfo {author} {\bibfnamefont {R.}~\bibnamefont
  {Lora-Serrano}}, \bibinfo {author} {\bibfnamefont {C.}~\bibnamefont {Giles}},
  \bibinfo {author} {\bibfnamefont {E.}~\bibnamefont {Granado}}, \bibinfo
  {author} {\bibfnamefont {D.~J.}\ \bibnamefont {Garcia}}, \bibinfo {author}
  {\bibfnamefont {E.}~\bibnamefont {Miranda}}, \bibinfo {author} {\bibfnamefont
  {O.}~\bibnamefont {Ag\"uero}}, \bibinfo {author} {\bibfnamefont
  {L.}~\bibnamefont {Mendon\ifmmode \mbox{\c{c}}\else~\c{c}\fi{}a Ferreira}},
  \bibinfo {author} {\bibfnamefont {J.~G.~S.}\ \bibnamefont {Duque}}, \ and\
  \bibinfo {author} {\bibfnamefont {P.~G.}\ \bibnamefont {Pagliuso}},\ }\href
  {\doibase 10.1103/PhysRevB.74.214404} {\bibfield  {journal} {\bibinfo
  {journal} {Phys. Rev. B}\ }\textbf {\bibinfo {volume} {74}},\ \bibinfo
  {pages} {214404} (\bibinfo {year} {2006})}\BibitemShut {NoStop}%
\bibitem [{\citenamefont {Van~Hieu}\ \emph {et~al.}(2006)\citenamefont
  {Van~Hieu}, \citenamefont {Shishido}, \citenamefont {Takeuchi}, \citenamefont
  {Thamizhavel}, \citenamefont {Nakashima}, \citenamefont {Sugiyama},
  \citenamefont {Settai}, \citenamefont {D.~Matsuda}, \citenamefont {Haga},
  \citenamefont {Hagiwara}, \citenamefont {Kindo},\ and\ \citenamefont
  {Onuki}}]{JPSJ.75.074708}%
  \BibitemOpen
  \bibfield  {author} {\bibinfo {author} {\bibfnamefont {N.}~\bibnamefont
  {Van~Hieu}}, \bibinfo {author} {\bibfnamefont {H.}~\bibnamefont {Shishido}},
  \bibinfo {author} {\bibfnamefont {T.}~\bibnamefont {Takeuchi}}, \bibinfo
  {author} {\bibfnamefont {A.}~\bibnamefont {Thamizhavel}}, \bibinfo {author}
  {\bibfnamefont {H.}~\bibnamefont {Nakashima}}, \bibinfo {author}
  {\bibfnamefont {K.}~\bibnamefont {Sugiyama}}, \bibinfo {author}
  {\bibfnamefont {R.}~\bibnamefont {Settai}}, \bibinfo {author} {\bibfnamefont
  {T.}~\bibnamefont {D.~Matsuda}}, \bibinfo {author} {\bibfnamefont
  {Y.}~\bibnamefont {Haga}}, \bibinfo {author} {\bibfnamefont {M.}~\bibnamefont
  {Hagiwara}}, \bibinfo {author} {\bibfnamefont {K.}~\bibnamefont {Kindo}}, \
  and\ \bibinfo {author} {\bibfnamefont {Y.}~\bibnamefont {Onuki}},\ }\href
  {\doibase 10.1143/JPSJ.75.074708} {\bibfield  {journal} {\bibinfo  {journal}
  {Journal of the Physical Society of Japan}\ }\textbf {\bibinfo {volume}
  {75}},\ \bibinfo {pages} {074708} (\bibinfo {year} {2006})},\ \Eprint
  {http://arxiv.org/abs/http://dx.doi.org/10.1143/JPSJ.75.074708}
  {http://dx.doi.org/10.1143/JPSJ.75.074708} \BibitemShut {NoStop}%
\bibitem [{\citenamefont {Petrovic}\ \emph
  {et~al.}(2001{\natexlab{b}})\citenamefont {Petrovic}, \citenamefont
  {Movshovich}, \citenamefont {Jaime}, \citenamefont {Pagliuso}, \citenamefont
  {Hundley}, \citenamefont {Sarrao}, \citenamefont {Fisk},\ and\ \citenamefont
  {Thompson}}]{0295-5075-53-3-354}%
  \BibitemOpen
  \bibfield  {author} {\bibinfo {author} {\bibfnamefont {C.}~\bibnamefont
  {Petrovic}}, \bibinfo {author} {\bibfnamefont {R.}~\bibnamefont
  {Movshovich}}, \bibinfo {author} {\bibfnamefont {M.}~\bibnamefont {Jaime}},
  \bibinfo {author} {\bibfnamefont {P.~G.}\ \bibnamefont {Pagliuso}}, \bibinfo
  {author} {\bibfnamefont {M.~F.}\ \bibnamefont {Hundley}}, \bibinfo {author}
  {\bibfnamefont {J.~L.}\ \bibnamefont {Sarrao}}, \bibinfo {author}
  {\bibfnamefont {Z.}~\bibnamefont {Fisk}}, \ and\ \bibinfo {author}
  {\bibfnamefont {J.~D.}\ \bibnamefont {Thompson}},\ }\href
  {http://stacks.iop.org/0295-5075/53/i=3/a=354} {\bibfield  {journal}
  {\bibinfo  {journal} {EPL (Europhysics Letters)}\ }\textbf {\bibinfo {volume}
  {53}},\ \bibinfo {pages} {354} (\bibinfo {year}
  {2001}{\natexlab{b}})}\BibitemShut {NoStop}%
\bibitem [{\citenamefont {Serrano}(2006)}]{serrano2006determinaccao}%
  \BibitemOpen
  \bibfield  {author} {\bibinfo {author} {\bibfnamefont {R.~L.}\ \bibnamefont
  {Serrano}},\ }\href@noop {} {\emph {\bibinfo {title} {Determina{\c{c}}ao de
  estruturas magn{\'e}ticas de novos compostos intermet{\'a}licos}}}\ (\bibinfo
   {publisher} {Biblioteca Digital da Unicamp},\ \bibinfo {year}
  {2006})\BibitemShut {NoStop}%
\bibitem [{\citenamefont {Anisimov}\ \emph {et~al.}(2005)\citenamefont
  {Anisimov}, \citenamefont {Kondakov}, \citenamefont {Kozhevnikov},
  \citenamefont {Nekrasov}, \citenamefont {Pchelkina}, \citenamefont {Allen},
  \citenamefont {Mo}, \citenamefont {Kim}, \citenamefont {Metcalf},
  \citenamefont {Suga}, \citenamefont {Sekiyama}, \citenamefont {Keller},
  \citenamefont {Leonov}, \citenamefont {Ren},\ and\ \citenamefont
  {Vollhardt}}]{Anisimov2005}%
  \BibitemOpen
  \bibfield  {author} {\bibinfo {author} {\bibfnamefont {V.~I.}\ \bibnamefont
  {Anisimov}}, \bibinfo {author} {\bibfnamefont {D.~E.}\ \bibnamefont
  {Kondakov}}, \bibinfo {author} {\bibfnamefont {A.~V.}\ \bibnamefont
  {Kozhevnikov}}, \bibinfo {author} {\bibfnamefont {I.~A.}\ \bibnamefont
  {Nekrasov}}, \bibinfo {author} {\bibfnamefont {Z.~V.}\ \bibnamefont
  {Pchelkina}}, \bibinfo {author} {\bibfnamefont {J.~W.}\ \bibnamefont
  {Allen}}, \bibinfo {author} {\bibfnamefont {S.-K.}\ \bibnamefont {Mo}},
  \bibinfo {author} {\bibfnamefont {H.-D.}\ \bibnamefont {Kim}}, \bibinfo
  {author} {\bibfnamefont {P.}~\bibnamefont {Metcalf}}, \bibinfo {author}
  {\bibfnamefont {S.}~\bibnamefont {Suga}}, \bibinfo {author} {\bibfnamefont
  {A.}~\bibnamefont {Sekiyama}}, \bibinfo {author} {\bibfnamefont
  {G.}~\bibnamefont {Keller}}, \bibinfo {author} {\bibfnamefont
  {I.}~\bibnamefont {Leonov}}, \bibinfo {author} {\bibfnamefont
  {X.}~\bibnamefont {Ren}}, \ and\ \bibinfo {author} {\bibfnamefont
  {D.}~\bibnamefont {Vollhardt}},\ }\href {\doibase 10.1103/PhysRevB.71.125119}
  {\bibfield  {journal} {\bibinfo  {journal} {Phys. Rev. B}\ }\textbf {\bibinfo
  {volume} {71}},\ \bibinfo {pages} {125119} (\bibinfo {year}
  {2005})}\BibitemShut {NoStop}%
\bibitem [{\citenamefont {Blaha}\ \emph {et~al.}(2001)\citenamefont {Blaha},
  \citenamefont {Schwarz}, \citenamefont {Madsen}, \citenamefont {Kvasnicka},\
  and\ \citenamefont {Luitz}}]{wien2k}%
  \BibitemOpen
  \bibfield  {author} {\bibinfo {author} {\bibfnamefont {P.}~\bibnamefont
  {Blaha}}, \bibinfo {author} {\bibfnamefont {K.}~\bibnamefont {Schwarz}},
  \bibinfo {author} {\bibfnamefont {G.~K.~H.}\ \bibnamefont {Madsen}}, \bibinfo
  {author} {\bibfnamefont {D.}~\bibnamefont {Kvasnicka}}, \ and\ \bibinfo
  {author} {\bibfnamefont {J.}~\bibnamefont {Luitz}},\ }\href@noop {} {\emph
  {\bibinfo {title} {{WIEN2K}, {A}n {A}ugmented {P}lane {W}ave + {L}ocal
  {O}rbitals {P}rogram for {C}alculating {C}rystal {P}roperties}}}\ (\bibinfo
  {publisher} {{K}arlheinz Schwarz, Techn. Universit\"{a}t Wien, Austria},\
  \bibinfo {year} {2001})\BibitemShut {NoStop}%
\bibitem [{\citenamefont {Perdew}\ \emph {et~al.}(1996)\citenamefont {Perdew},
  \citenamefont {Burke},\ and\ \citenamefont {Ernzerhof}}]{Perdew1996}%
  \BibitemOpen
  \bibfield  {author} {\bibinfo {author} {\bibfnamefont {J.~P.}\ \bibnamefont
  {Perdew}}, \bibinfo {author} {\bibfnamefont {K.}~\bibnamefont {Burke}}, \
  and\ \bibinfo {author} {\bibfnamefont {M.}~\bibnamefont {Ernzerhof}},\
  }\href@noop {} {\ ,\ \bibinfo {pages} {3865} (\bibinfo {year}
  {1996})}\BibitemShut {NoStop}%
\bibitem [{\citenamefont {Piekarz}\ \emph {et~al.}(2005)\citenamefont
  {Piekarz}, \citenamefont {Parlinski}, \citenamefont {Jochym}, \citenamefont
  {Sanchez},\ and\ \citenamefont {Rebizant}}]{Piekarz2005}%
  \BibitemOpen
  \bibfield  {author} {\bibinfo {author} {\bibfnamefont {P.}~\bibnamefont
  {Piekarz}}, \bibinfo {author} {\bibfnamefont {K.}~\bibnamefont {Parlinski}},
  \bibinfo {author} {\bibfnamefont {P.~T.}\ \bibnamefont {Jochym}}, \bibinfo
  {author} {\bibfnamefont {J.-P.}\ \bibnamefont {Sanchez}}, \ and\ \bibinfo
  {author} {\bibfnamefont {J.}~\bibnamefont {Rebizant}},\ }\href {\doibase
  10.1103/PhysRevB.72.014521} {\bibfield  {journal} {\bibinfo  {journal} {Phys.
  Rev. B}\ }\textbf {\bibinfo {volume} {72}},\ \bibinfo {pages} {014521}
  (\bibinfo {year} {2005})}\BibitemShut {NoStop}%
\bibitem [{\citenamefont {Raymond}\ \emph {et~al.}(2006)\citenamefont
  {Raymond}, \citenamefont {Piekarz}, \citenamefont {Sanchez}, \citenamefont
  {Serrano}, \citenamefont {Krisch}, \citenamefont
  {Janou\ifmmode~\check{s}\else \v{s}\fi{}ov\'a}, \citenamefont {Rebizant},
  \citenamefont {Metoki}, \citenamefont {Kaneko}, \citenamefont {Jochym},
  \citenamefont {Ole\ifmmode~\acute{s}\else \'{s}\fi{}},\ and\ \citenamefont
  {Parlinski}}]{PhysRevLett.96.237003}%
  \BibitemOpen
  \bibfield  {author} {\bibinfo {author} {\bibfnamefont {S.}~\bibnamefont
  {Raymond}}, \bibinfo {author} {\bibfnamefont {P.}~\bibnamefont {Piekarz}},
  \bibinfo {author} {\bibfnamefont {J.~P.}\ \bibnamefont {Sanchez}}, \bibinfo
  {author} {\bibfnamefont {J.}~\bibnamefont {Serrano}}, \bibinfo {author}
  {\bibfnamefont {M.}~\bibnamefont {Krisch}}, \bibinfo {author} {\bibfnamefont
  {B.}~\bibnamefont {Janou\ifmmode~\check{s}\else \v{s}\fi{}ov\'a}}, \bibinfo
  {author} {\bibfnamefont {J.}~\bibnamefont {Rebizant}}, \bibinfo {author}
  {\bibfnamefont {N.}~\bibnamefont {Metoki}}, \bibinfo {author} {\bibfnamefont
  {K.}~\bibnamefont {Kaneko}}, \bibinfo {author} {\bibfnamefont {P.~T.}\
  \bibnamefont {Jochym}}, \bibinfo {author} {\bibfnamefont {A.~M.}\
  \bibnamefont {Ole\ifmmode~\acute{s}\else \'{s}\fi{}}}, \ and\ \bibinfo
  {author} {\bibfnamefont {K.}~\bibnamefont {Parlinski}},\ }\href {\doibase
  10.1103/PhysRevLett.96.237003} {\bibfield  {journal} {\bibinfo  {journal}
  {Phys. Rev. Lett.}\ }\textbf {\bibinfo {volume} {96}},\ \bibinfo {pages}
  {237003} (\bibinfo {year} {2006})}\BibitemShut {NoStop}%
\bibitem [{\citenamefont {Zhu}\ \emph {et~al.}(2012)\citenamefont {Zhu},
  \citenamefont {Tobash}, \citenamefont {Bauer}, \citenamefont {Ronning},
  \citenamefont {Scott}, \citenamefont {Haule}, \citenamefont {Kotliar},
  \citenamefont {Albers},\ and\ \citenamefont {Wills}}]{Zhu2012}%
  \BibitemOpen
  \bibfield  {author} {\bibinfo {author} {\bibfnamefont {J.-X.}\ \bibnamefont
  {Zhu}}, \bibinfo {author} {\bibfnamefont {P.~H.}\ \bibnamefont {Tobash}},
  \bibinfo {author} {\bibfnamefont {E.~D.}\ \bibnamefont {Bauer}}, \bibinfo
  {author} {\bibfnamefont {F.}~\bibnamefont {Ronning}}, \bibinfo {author}
  {\bibfnamefont {B.~L.}\ \bibnamefont {Scott}}, \bibinfo {author}
  {\bibfnamefont {K.}~\bibnamefont {Haule}}, \bibinfo {author} {\bibfnamefont
  {G.}~\bibnamefont {Kotliar}}, \bibinfo {author} {\bibfnamefont {R.~C.}\
  \bibnamefont {Albers}}, \ and\ \bibinfo {author} {\bibfnamefont {J.~M.}\
  \bibnamefont {Wills}},\ }\href {\doibase 10.1209/0295-5075/97/57001}
  {\bibfield  {journal} {\bibinfo  {journal} {EPL}\ }\textbf {\bibinfo {volume}
  {97}},\ \bibinfo {pages} {57001} (\bibinfo {year} {2012})}\BibitemShut
  {NoStop}%
\bibitem [{\citenamefont {Anisimov}\ \emph {et~al.}(1993)\citenamefont
  {Anisimov}, \citenamefont {Solovyev}, \citenamefont {Korotin}, \citenamefont
  {Czy\ifmmode~\dot{z}\else \.{z}\fi{}yk},\ and\ \citenamefont
  {Sawatzky}}]{Anisimov1993}%
  \BibitemOpen
  \bibfield  {author} {\bibinfo {author} {\bibfnamefont {V.~I.}\ \bibnamefont
  {Anisimov}}, \bibinfo {author} {\bibfnamefont {I.~V.}\ \bibnamefont
  {Solovyev}}, \bibinfo {author} {\bibfnamefont {M.~A.}\ \bibnamefont
  {Korotin}}, \bibinfo {author} {\bibfnamefont {M.~T.}\ \bibnamefont
  {Czy\ifmmode~\dot{z}\else \.{z}\fi{}yk}}, \ and\ \bibinfo {author}
  {\bibfnamefont {G.~A.}\ \bibnamefont {Sawatzky}},\ }\href {\doibase
  10.1103/PhysRevB.48.16929} {\bibfield  {journal} {\bibinfo  {journal} {Phys.
  Rev. B}\ }\textbf {\bibinfo {volume} {48}},\ \bibinfo {pages} {16929}
  (\bibinfo {year} {1993})}\BibitemShut {NoStop}%
\bibitem [{\citenamefont {Yin}\ and\ \citenamefont {Pickett}(2006)}]{Yin2006}%
  \BibitemOpen
  \bibfield  {author} {\bibinfo {author} {\bibfnamefont {Z.}~\bibnamefont
  {Yin}}\ and\ \bibinfo {author} {\bibfnamefont {W.}~\bibnamefont {Pickett}},\
  }\href {\doibase 10.1103/PhysRevB.74.205106} {\bibfield  {journal} {\bibinfo
  {journal} {Phys. Rev. B}\ }\textbf {\bibinfo {volume} {74}},\ \bibinfo
  {pages} {205106} (\bibinfo {year} {2006})}\BibitemShut {NoStop}%
\bibitem [{\citenamefont {Petersen}\ \emph {et~al.}(2006)\citenamefont
  {Petersen}, \citenamefont {Hafner},\ and\ \citenamefont
  {Marsman}}]{Petersen2006}%
  \BibitemOpen
  \bibfield  {author} {\bibinfo {author} {\bibfnamefont {M.}~\bibnamefont
  {Petersen}}, \bibinfo {author} {\bibfnamefont {J.}~\bibnamefont {Hafner}}, \
  and\ \bibinfo {author} {\bibfnamefont {M.}~\bibnamefont {Marsman}},\ }\href
  {http://stacks.iop.org/0953-8984/18/i=30/a=007} {\bibfield  {journal}
  {\bibinfo  {journal} {Journal of Physics: Condensed Matter}\ }\textbf
  {\bibinfo {volume} {18}},\ \bibinfo {pages} {7021} (\bibinfo {year}
  {2006})}\BibitemShut {NoStop}%
\bibitem [{\citenamefont {Granado}\ \emph {et~al.}(2006)\citenamefont
  {Granado}, \citenamefont {Uchoa}, \citenamefont {Malachias}, \citenamefont
  {Lora-Serrano}, \citenamefont {Pagliuso},\ and\ \citenamefont
  {Westfahl}}]{Granado2006}%
  \BibitemOpen
  \bibfield  {author} {\bibinfo {author} {\bibfnamefont {E.}~\bibnamefont
  {Granado}}, \bibinfo {author} {\bibfnamefont {B.}~\bibnamefont {Uchoa}},
  \bibinfo {author} {\bibfnamefont {a.}~\bibnamefont {Malachias}}, \bibinfo
  {author} {\bibfnamefont {R.}~\bibnamefont {Lora-Serrano}}, \bibinfo {author}
  {\bibfnamefont {P.}~\bibnamefont {Pagliuso}}, \ and\ \bibinfo {author}
  {\bibfnamefont {H.}~\bibnamefont {Westfahl}},\ }\href {\doibase
  10.1103/PhysRevB.74.214428} {\bibfield  {journal} {\bibinfo  {journal} {Phys.
  Rev. B}\ }\textbf {\bibinfo {volume} {74}},\ \bibinfo {pages} {214428}
  (\bibinfo {year} {2006})}\BibitemShut {NoStop}%
\bibitem [{\citenamefont {Chang}\ \emph {et~al.}(2002)\citenamefont {Chang},
  \citenamefont {Pagliuso}, \citenamefont {Bao}, \citenamefont {Gardner},
  \citenamefont {Swainson}, \citenamefont {Sarrao},\ and\ \citenamefont
  {Nakotte}}]{Chang2002}%
  \BibitemOpen
  \bibfield  {author} {\bibinfo {author} {\bibfnamefont {S.}~\bibnamefont
  {Chang}}, \bibinfo {author} {\bibfnamefont {P.~G.}\ \bibnamefont {Pagliuso}},
  \bibinfo {author} {\bibfnamefont {W.}~\bibnamefont {Bao}}, \bibinfo {author}
  {\bibfnamefont {J.~S.}\ \bibnamefont {Gardner}}, \bibinfo {author}
  {\bibfnamefont {I.~P.}\ \bibnamefont {Swainson}}, \bibinfo {author}
  {\bibfnamefont {J.~L.}\ \bibnamefont {Sarrao}}, \ and\ \bibinfo {author}
  {\bibfnamefont {H.}~\bibnamefont {Nakotte}},\ }\href {\doibase
  10.1103/PhysRevB.66.132417} {\bibfield  {journal} {\bibinfo  {journal} {Phys.
  Rev. B}\ }\textbf {\bibinfo {volume} {66}},\ \bibinfo {pages} {132417}
  (\bibinfo {year} {2002})}\BibitemShut {NoStop}%
\bibitem [{\citenamefont {Hieu}\ \emph {et~al.}(2006)\citenamefont {Hieu},
  \citenamefont {Shishido}, \citenamefont {Takeuchi}, \citenamefont
  {Thamizhavel}, \citenamefont {Nakashima}, \citenamefont {Sugiyama},
  \citenamefont {Settai}, \citenamefont {Matsuda}, \citenamefont {Haga},
  \citenamefont {Hagiwara}, \citenamefont {Kindo},\ and\ \citenamefont
  {\=Onuki}}]{vanHieu1}%
  \BibitemOpen
  \bibfield  {author} {\bibinfo {author} {\bibfnamefont {N.~V.}\ \bibnamefont
  {Hieu}}, \bibinfo {author} {\bibfnamefont {H.}~\bibnamefont {Shishido}},
  \bibinfo {author} {\bibfnamefont {T.}~\bibnamefont {Takeuchi}}, \bibinfo
  {author} {\bibfnamefont {A.}~\bibnamefont {Thamizhavel}}, \bibinfo {author}
  {\bibfnamefont {H.}~\bibnamefont {Nakashima}}, \bibinfo {author}
  {\bibfnamefont {K.}~\bibnamefont {Sugiyama}}, \bibinfo {author}
  {\bibfnamefont {R.}~\bibnamefont {Settai}}, \bibinfo {author} {\bibfnamefont
  {T.~D.}\ \bibnamefont {Matsuda}}, \bibinfo {author} {\bibfnamefont
  {Y.}~\bibnamefont {Haga}}, \bibinfo {author} {\bibfnamefont {M.}~\bibnamefont
  {Hagiwara}}, \bibinfo {author} {\bibfnamefont {K.}~\bibnamefont {Kindo}}, \
  and\ \bibinfo {author} {\bibfnamefont {Y.}~\bibnamefont {\=Onuki}},\
  }\href@noop {} {\bibfield  {journal} {\bibinfo  {journal} {J.\ Phys. Soc.
  Jpn.}\ }\textbf {\bibinfo {volume} {75}},\ \bibinfo {pages} {074708}
  (\bibinfo {year} {2006})}\BibitemShut {NoStop}%
\bibitem [{Jen()}]{Jensen&Mackintosh}%
  \BibitemOpen
  \href@noop {} {}\bibinfo {note} {Rare Earth Magnetism Structures and
  Excitations, CLARENDON PRESS - OXFORD 1991.}\BibitemShut {Stop}%
\bibitem [{\citenamefont {Cabrera-Baez}\ \emph {et~al.}(2014)\citenamefont
  {Cabrera-Baez}, \citenamefont {Iwamoto}, \citenamefont {Magnavita},
  \citenamefont {Osorio-Guill\'{e}n}, \citenamefont {Ribeiro}, \citenamefont
  {Avila},\ and\ \citenamefont {Rettori}}]{Cabrera-Baez2014}%
  \BibitemOpen
  \bibfield  {author} {\bibinfo {author} {\bibfnamefont {M.}~\bibnamefont
  {Cabrera-Baez}}, \bibinfo {author} {\bibfnamefont {W.}~\bibnamefont
  {Iwamoto}}, \bibinfo {author} {\bibfnamefont {E.~T.}\ \bibnamefont
  {Magnavita}}, \bibinfo {author} {\bibfnamefont {J.~M.}\ \bibnamefont
  {Osorio-Guill\'{e}n}}, \bibinfo {author} {\bibfnamefont {R.~a.}\ \bibnamefont
  {Ribeiro}}, \bibinfo {author} {\bibfnamefont {M.~a.}\ \bibnamefont {Avila}},
  \ and\ \bibinfo {author} {\bibfnamefont {C.}~\bibnamefont {Rettori}},\ }\href
  {\doibase 10.1088/0953-8984/26/17/175501} {\bibfield  {journal} {\bibinfo
  {journal} {J. Phys.: Condens. Matter}\ }\textbf {\bibinfo {volume} {26}},\
  \bibinfo {pages} {175501} (\bibinfo {year} {2014})}\BibitemShut {NoStop}%
\bibitem [{\citenamefont {Levy}\ \emph {et~al.}(2012)\citenamefont {Levy},
  \citenamefont {Sutarto}, \citenamefont {Chevrier}, \citenamefont {Regier},
  \citenamefont {Blyth}, \citenamefont {Geck}, \citenamefont {Wurmehl},
  \citenamefont {Harnagea}, \citenamefont {Wadati}, \citenamefont {Mizokawa},
  \citenamefont {Elfimov}, \citenamefont {Damascelli},\ and\ \citenamefont
  {Sawatzky}}]{Levy2012}%
  \BibitemOpen
  \bibfield  {author} {\bibinfo {author} {\bibfnamefont {G.}~\bibnamefont
  {Levy}}, \bibinfo {author} {\bibfnamefont {R.}~\bibnamefont {Sutarto}},
  \bibinfo {author} {\bibfnamefont {D.}~\bibnamefont {Chevrier}}, \bibinfo
  {author} {\bibfnamefont {T.}~\bibnamefont {Regier}}, \bibinfo {author}
  {\bibfnamefont {R.}~\bibnamefont {Blyth}}, \bibinfo {author} {\bibfnamefont
  {J.}~\bibnamefont {Geck}}, \bibinfo {author} {\bibfnamefont {S.}~\bibnamefont
  {Wurmehl}}, \bibinfo {author} {\bibfnamefont {L.}~\bibnamefont {Harnagea}},
  \bibinfo {author} {\bibfnamefont {H.}~\bibnamefont {Wadati}}, \bibinfo
  {author} {\bibfnamefont {T.}~\bibnamefont {Mizokawa}}, \bibinfo {author}
  {\bibfnamefont {I.~S.}\ \bibnamefont {Elfimov}}, \bibinfo {author}
  {\bibfnamefont {A.}~\bibnamefont {Damascelli}}, \ and\ \bibinfo {author}
  {\bibfnamefont {G.~A.}\ \bibnamefont {Sawatzky}},\ }\href {\doibase
  10.1103/PhysRevLett.109.077001} {\bibfield  {journal} {\bibinfo  {journal}
  {Phys. Rev. Lett.}\ }\textbf {\bibinfo {volume} {109}},\ \bibinfo {pages}
  {077001} (\bibinfo {year} {2012})}\BibitemShut {NoStop}%
\bibitem [{\citenamefont {Blanco}\ \emph {et~al.}(1991)\citenamefont {Blanco},
  \citenamefont {Gignoux},\ and\ \citenamefont {Schmitt}}]{Blanco1991}%
  \BibitemOpen
  \bibfield  {author} {\bibinfo {author} {\bibfnamefont {J.}~\bibnamefont
  {Blanco}}, \bibinfo {author} {\bibfnamefont {D.}~\bibnamefont {Gignoux}}, \
  and\ \bibinfo {author} {\bibfnamefont {D.}~\bibnamefont {Schmitt}},\ }\href
  {http://prb.aps.org/abstract/PRB/v43/i16/p13145\_1} {\bibfield  {journal}
  {\bibinfo  {journal} {Phys. Rev. B}\ }\textbf {\bibinfo {volume} {43}},\
  \bibinfo {pages} {1} (\bibinfo {year} {1991})}\BibitemShut {NoStop}%
\bibitem [{\citenamefont {Sandvik}(1999)}]{alps}%
  \BibitemOpen
  \bibfield  {author} {\bibinfo {author} {\bibfnamefont {A.~W.}\ \bibnamefont
  {Sandvik}},\ }\href@noop {} {\bibfield  {journal} {\bibinfo  {journal} {Phys.
  Rev. B}\ }\textbf {\bibinfo {volume} {59}},\ \bibinfo {pages} {14157}
  (\bibinfo {year} {1999})}\BibitemShut {NoStop}%
\bibitem [{Note1()}]{Note1}%
  \BibitemOpen
  \bibinfo {note} {A detailed analysis of the phonon contribution to the
  specific heat in these compounds will be presented elsewhere.}\BibitemShut
  {Stop}%
\bibitem [{\citenamefont {Wallace}(1998)}]{Wallace}%
  \BibitemOpen
  \bibfield  {author} {\bibinfo {author} {\bibfnamefont {D.}~\bibnamefont
  {Wallace}},\ }\href {http://books.google.se/books?id=qLzOmwSgMIsC} {\emph
  {\bibinfo {title} {Thermodynamics of Crystals}}},\ Dover books on physics\
  (\bibinfo  {publisher} {Dover Publications},\ \bibinfo {year}
  {1998})\BibitemShut {NoStop}%
\bibitem [{\citenamefont {Oitmaa}\ and\ \citenamefont
  {Zheng}(2004)}]{Oitmaa2004}%
  \BibitemOpen
  \bibfield  {author} {\bibinfo {author} {\bibfnamefont {J.}~\bibnamefont
  {Oitmaa}}\ and\ \bibinfo {author} {\bibfnamefont {W.}~\bibnamefont {Zheng}},\
  }\href {http://stacks.iop.org/0953-8984/16/i=47/a=016} {\bibfield  {journal}
  {\bibinfo  {journal} {J. Phys: Condens. Matter}\ }\textbf {\bibinfo {volume}
  {16}},\ \bibinfo {pages} {8653} (\bibinfo {year} {2004})}\BibitemShut
  {NoStop}%
\end{thebibliography}%
\bibliographystyle{apsrev4-1}
\end{document}